\documentclass[useAMS,usenatbib]{mn2e}
\usepackage{mathtools}
\usepackage{pdflscape}
\usepackage{longtable}
\usepackage{graphicx}
\usepackage{natbib}
\usepackage{lscape}
\usepackage{longtable}
\usepackage{subfig}
\usepackage{amssymb,amsmath}
\usepackage[T1]{fontenc}
\usepackage{hyperref}
\usepackage{amsmath}
\usepackage{breakurl}
\usepackage{times}

\newcommand{\CIV}{C\,{\sc iv}}
\newcommand{\MgII}{Mg\,{\sc ii}}

\newcommand{\CIII}{C\,{\sc iii}]}

\newcommand{\SiIV}{Si\,{\sc iv}}

\newcommand\ion[2]{#1$\;${\small\rmfamily\@Roman{#2}}\relax}%

\def\lsim{\lower0.3em\hbox{$\,\buildrel <\over\sim\,$}}
\def\gsim{\lower0.3em\hbox{$\,\buildrel >\over\sim\,$}}

\title[Quasar Similarity]{How Similar are the Properties of Quasars with Nearly Identical Ultraviolet Spectra?}
\author[Rochais, Singh, Chick, Maithil, Sutter, Brotherton \& Shang]{Thomas Rochais$^{1}$, Vikram Singh$^{1}$, William Chick$^{1}$, Jaya Maithil$^{1}$, Jessica Sutter$^{1}$, \newauthor Michael S. Brotherton$^{1}$\thanks{E-mail: mbrother@uwyo.edu}, Zhaohui Shang$^{2}$ \\
$^{1}$Department of Physics and Astronomy, University of Wyoming, Laramie, WY 82071, USA\\ 
$^{2}$Department of Astronomy, Tianjin Normal University, China\\
}

\begin{document}
\date{}

\pagerange{\pageref{firstpage}--\pageref{lastpage}} \pubyear{2016}

\maketitle

\label{firstpage}

\begin{abstract}
The spectrum of a quasar contains important information about its properties. Thus, it can be expected that two quasars with similar spectra will have similar properties, but just how similar has not before been quantified. Here we compare the ultraviolet spectra of a sample of 5553 quasars from Data Release 7 of the Sloan Digital Sky Survey, focusing on the $1350$ \AA \ $\leq \lambda \leq 2900$ \AA \ rest-frame region which contains prominent emission lines from \SiIV, O IV], \CIV, \CIII, and \MgII\ species. We use principal component analysis to determine the dominant components of spectral variation, as well as to quantitatively measure spectral similarity. As suggested by both the Baldwin effect and modified Baldwin effect, quasars with similar spectra have similar properties:  bolometric luminosity, Eddington fraction, and black hole mass. 
The latter two quantities are calculated from the luminosity in conjunction with spectral features,
and the variation between quasars with virtually identical spectra (which we call doppelgangers)
is driven by the variance in the luminosity plus measurement uncertainties.
In the doppelgangers the luminosity differences show 1$\sigma$ uncertainties of 57\% (or 0.63 magnitudes) and $\sim$70\% 1$\sigma$ uncertainties for mass and Eddington fraction.
Much of the difference in luminosities may be attributable to time lags between the spectral lines
and the continuum.
Furthermore, we find that suggestions that the mostly highly accreting quasars should be better standard candles than other quasars are not bourne out for doppelgangers.  Finally, we discuss the implications for using quasars as cosmological probes and the nature of the first two spectral principal components.
\end{abstract}


\section{INTRODUCTION}

Baldwin (1977) reported the inverse correlation between the equivalent width (EW) of the C IV $\lambda$1549 emission line and the continuum emission.  This empirical relationship, now known as the Baldwin effect, was of interest for several reasons.  The fact that the line emission does not scale linearly with the underlying continuum suggests intriguing astrophysics (e.g., Mushotsky \& Ferland 1984; Netzer 1985; Wandel 1999).  More practically, a way to predict the continuum luminosity independent of redshift opened up the possibility of calibrating quasars as cosmological probes (e.g., Baldwin et al. 1978; Wampler et al. 1984).  The scatter in the Baldwin effect has unfortunately proven too large to let the relationship be of practical use for cosmology, at least with existing samples despite their now impressive size (e.g., Bian et al. 2012).

There are other correlations of interest relating quasar spectra and their physical properties.  For instance, the modified Baldwin effect is an inverse correlation between EW C IV and the Eddington fraction, L$_{bol}$/L$_{Edd}$ (e.g., Baskin \& Laor 2004; Shemmer et al. 2008; Shemmer \& Lieber 2015), generally stronger and more significant than the Baldwin effect itself.  Quasar masses are routinely estimated from single-epoch spectra using the velocity width of broad emission lines in combination with the continuum luminosity (Vestergaard \& Peterson 2006; Vestergaard \& Osmer 2009, and others).  Other parameters also leave their mark on the ultraviolet quasar spectrum, such as the orientation to the line of sight (e.g., Vestergaard 2002; Runnoe et al. 2014) and metallicity (e.g., Hamman \& Ferland 1999).  We must also be mindful of other issues, such as line-of-sight gas and dust that can redden spectra, or leave the imprints of absorption lines, which can complicate the measurement and interpretation of spectra.

Theoretically, the emitted quasar continuum and emission-line spectrum critically depend on parameters like the luminosity, which is related to the spectral energy distribution and ionizing continuum (e.g., Just et al. 2007).  The central black hole mass drives the kinematics of the broad-line region or BLR and the resulting observed line profiles (Peterson \& Wandel 1999).  The Eddington fraction depends on both the luminosity and the mass.  There are other factors, such as the specific geometries, quantities, and physical conditions of the line-emitting gas that are also important and may have significant variance from quasar to quasar, even if their underlying fundamental properties like black hole mass, accretion rate, and luminosity are the same.  It would be useful to quantify this variance statistically and learn to what extent an ultraviolet spectrum can be used to predict general quasar properties.

Traditionally, through the Baldwin effect and modified Baldwin effect, crude measurements like EW C IV have been used to predict luminosity and the Eddington fraction.  Modern computing allows spectra to be characterized in much more sophisticated ways than the equivalent width of a single line, and large data sets like the Sloan Digital Sky Survey (SDSS) and its quasar catalogs (e.g., Schneider et al. 2010) permit much more varied and extensive statistical comparisons.  There exist quasars with essentially identical ultraviolet spectra, to within differences associated with noise and absorption features, which we will refer to as doppelgangers.  We ask the question: how similar are the physical properties of quasar doppelgangers with nearly identical ultraviolet spectra?

We explicitly note that the black hole mass and the Eddington fraction are calculated using the continuum luminosity and the velocity widths of broad emission lines.  Therefore, for doppelganger pairs, any observed differences in these properties can be attributed to differences in luminosity plus spectral measurement uncertainties.  Luminosity then is the observable parameter driving all the differences in doppelganger pair properties.  We will consider each quasar property empirically but interpret our results in a self-consistent way in which luminosity variation is primary.  

In this paper, we use the technique of spectral principal component analysis (SPCA) to reconstruct quasar spectra without significant noise or absorption features, and to use their component weights to characterize how similar different spectra actually are.  In \S 2, we describe our sample selection, data, and SPCA approach.  In \S 3, we describe our analyses and results, examining how well very similar ultraviolet spectra can on average predict luminosity, the Eddington fraction L$_{bol}$/L$_{Edd}$, and black hole mass in quasars, as well as the special case of high Eddington fraction objects, which have been proposed to have more similar luminosities, thus potentially making them usable as cosmological probes.  We furthermore discuss our first two spectral principal components, which dominate the object-to-object spectral variation, what they physically represent, and how they correlate with quasar properties.  Finally, in \S 4 we discuss the implications of our results for cosmology and quasar astrophysics, along with outstanding sources of scatter, and in \S 5 summarize our general conclusions.  Throughout this paper, in order to be consistent with Shen et al. (2011), hereafter S11, we use cosmological parameters $\Omega_{\Lambda}$ = 0.7,  $\Omega_{m}$ = 0.3, and H$_{0}$=70 km s$^{-1}$ Mpc$^{-1}$.

\section{Sample, Data, and Methods}

\subsection{Sample and data}
We constructed our sample from the SDSS Data Release 7 (DR7) using measurements from S11. We selected quasars with the following criteria:
\begin{enumerate}
\item{Redshift range of $1.8 \leq z \leq 2.17$, placing the maximum number of useful diagnostic emission lines within the SDSS observed-frame spectral range.  On the short wavelength end is the $\lambda$1400 feature (a blend of Si IV and O IV] emission lines), and on the long wavelength end is Mg II $\lambda$2800, the best ultraviolet line for mass estimates (e.g., Trakhtenbrot \& Netzer 2012; Shen \& Liu 2012).  Also of interest are the C IV emission line and the blend of lines associated with C III] $\lambda$1909.  The relatively small redshift range also minimizes  potential cosmological issues.}
\item{No broad absorption lines (BALs) as indicated by the BAL flag = 0.  Broad absorption lines have strong effects on the observed spectrum, can account for significant variance, and make it impossible to evaluate the similarity of the unabsorbed emission.  Fortunately BAL quasars are a minority of the total population, although not totally unbiased with respect to physical properties like Eddington fraction (e.g., Ganguly et al. 2007).}
\item{Signal-to-noise ratio (SNR) greater than 10 in the C IV region.  While requiring higher SNR does limit the luminosity range somewhat, it becomes increasingly difficult to judge how similar quasar spectra are at lower SNR.  
}
\item{Measurements of observables of interest, including the rest-frame EW, FWHM, and velocity offset of C IV, and the BH mass based on \MgII\ (Vestergaard \& Osmer 2009, hereafter VO09).}
\end{enumerate}

We retrieved the Galactic-dereddened SDSS spectra of the resulting 5553 quasars identified using the above criteria from the website http://das.sdss.org/va/qso\_properties\_dr7/data/dered\_spectra/. Once downloaded, we used the SDSS pipeline redshift to put the spectra into their rest-frame and also resampled to 1 \AA\ bins spanning a common wavelength range from 1350 \AA\ to 2900 \AA.

\begin{figure}
\centering
\hspace{0cm}
   \includegraphics[width=9.5cm,trim={4cm 0 0 0},clip]{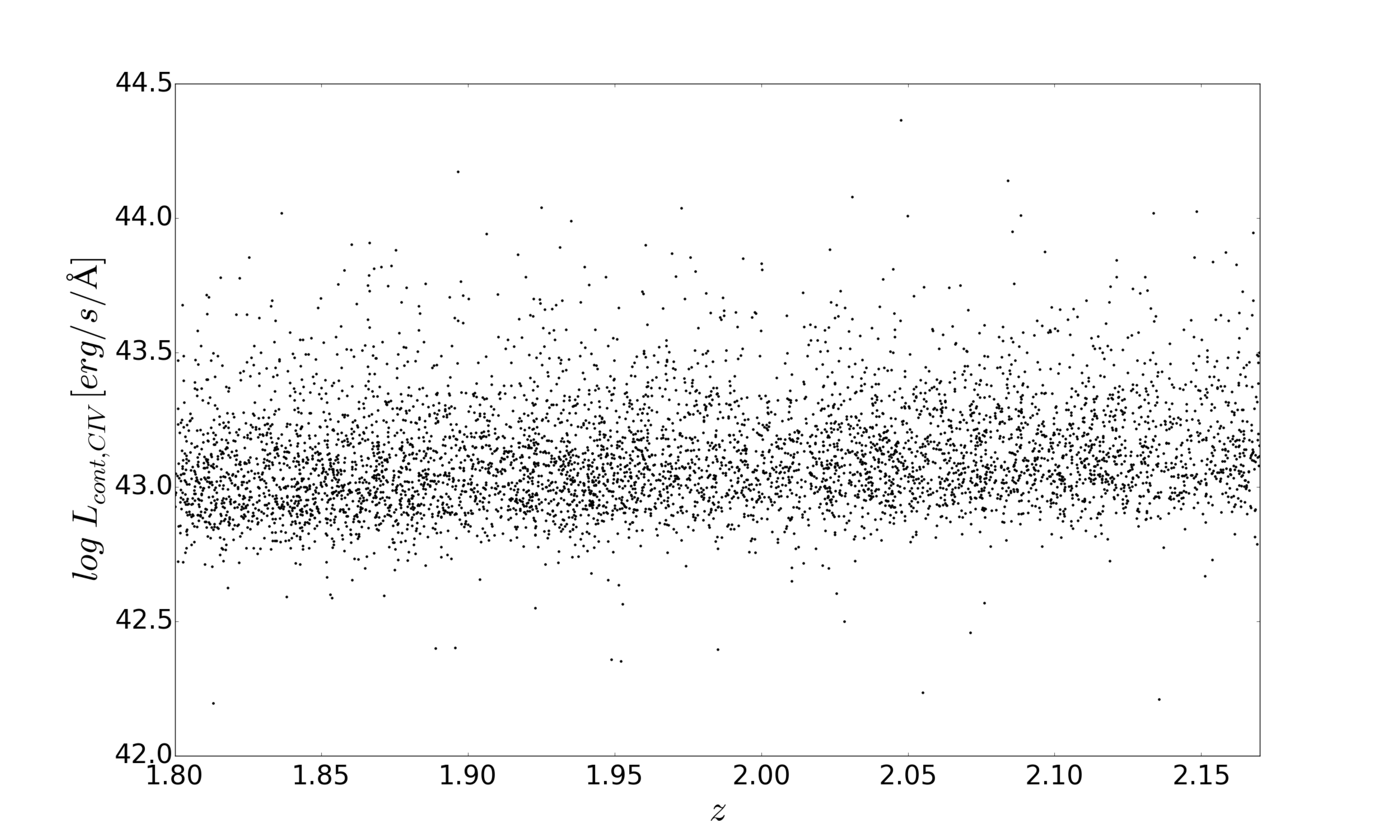}
   \vspace{0cm}
   \caption{Our sample in log luminosity vs. redshift space. Here the luminosity is the continuum luminosity at \CIV\ ($L_{cont,CIV}$) as defined in equation 1.  The mean error based on quoted uncertainies on luminosities in S11 for our sample is $\sim$0.01 dex.  Systematic uncertainties, for example those associated with orientation effects, are very likely dominant.}
   \label{fig:Lz}
\end{figure}
\begin{figure}
\centering
\hspace{0cm}
   \includegraphics[width=9.5cm,trim={4cm 0 0 0},clip]{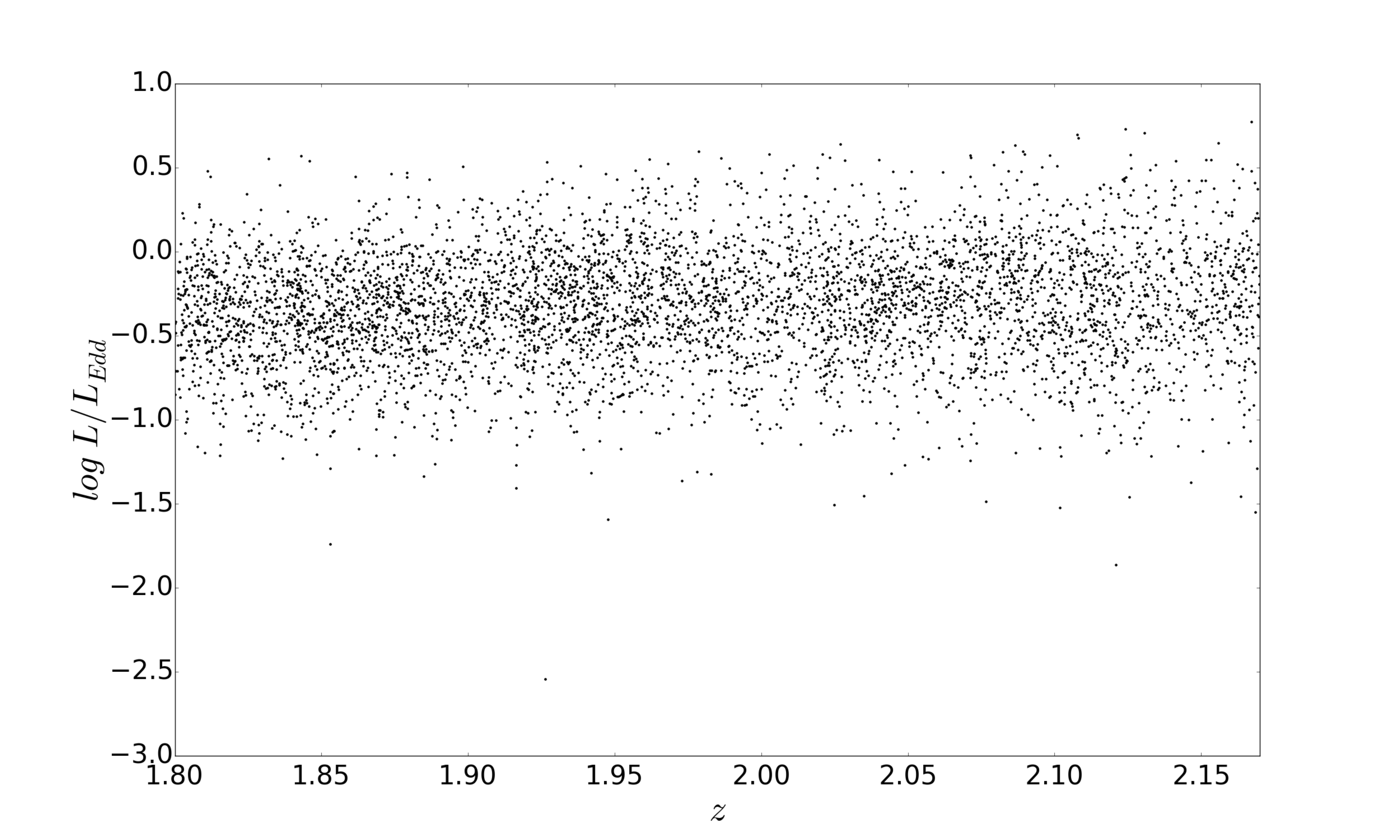}
   \vspace{0cm}
   \caption{Our sample in log Eddington fraction vs. redshift space. Here the Eddington fraction is the one computed using the S11 L$_{Bol}$ luminosity and L$_{Edd}$ based on Mg II BH mass as defined in equation 2.  Measurement uncertainties are dominated by the mass term, and are on average 0.16 dex for our sample, but the total uncertainty is likely dominated by systematic errors.}
   \label{fig:Eddz}
\end{figure}
\begin{figure}
\centering
\hspace{0cm}
   \includegraphics[width=9.5cm,trim={4cm 0 0 0},clip]{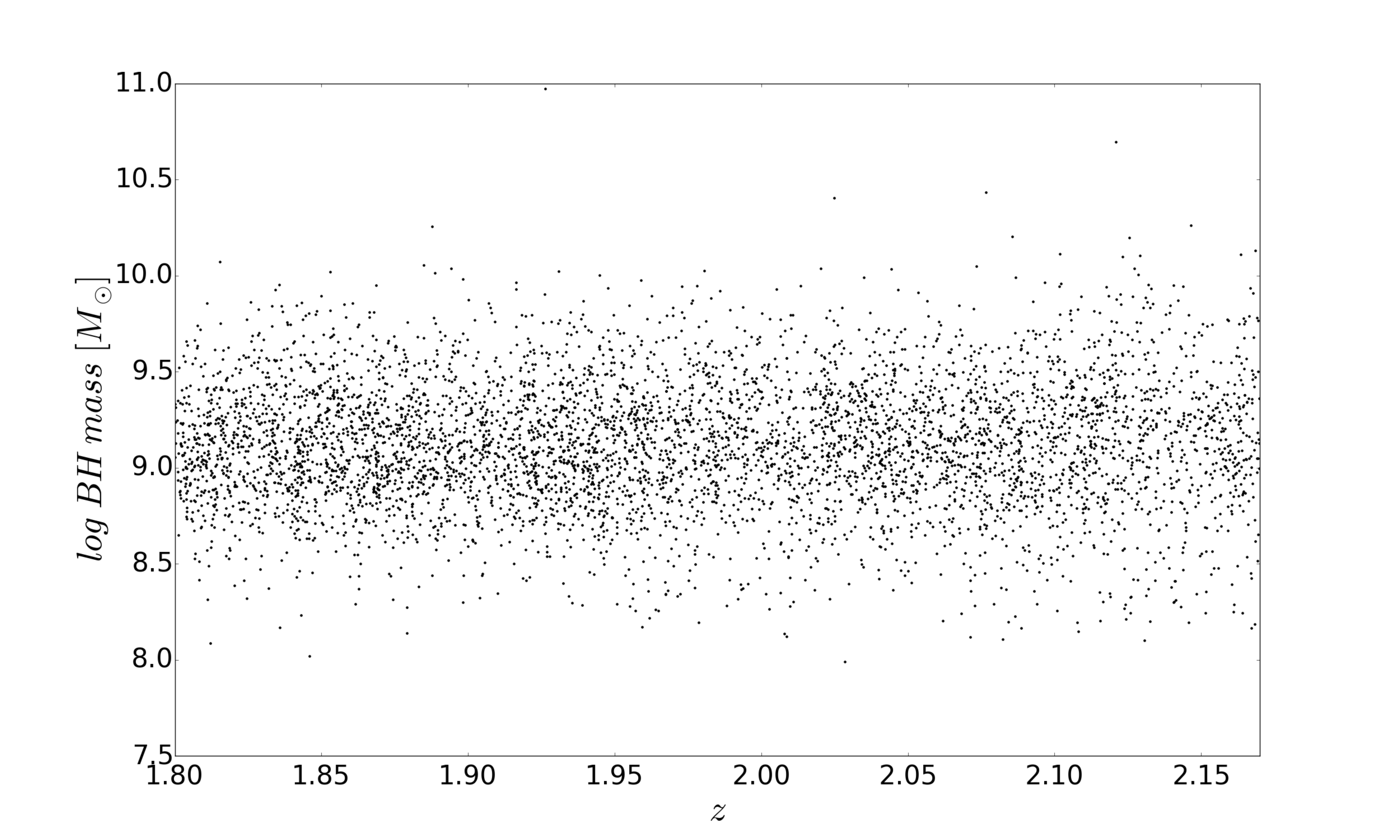}
   \vspace{0cm}
   \caption{Our sample in log BH mass vs. redshift space. Here the mass used is the BH mass based on \MgII \ VO09. Measurement uncertainties are on average 0.16 dex for our sample based on values reported by S11.  Systematic uncertainties on the actual mass are very likely larger and dominant.}
   \label{fig:Mz}
\end{figure}

We also modified or added several measurements to those of S11.
We computed the continuum luminosity at \CIV \ as:
\begin{equation}
L_{cont,CIV} = L_{\textrm{\CIV}}/EW_{\textrm{\CIV}}
\end{equation}
In practice this correlates very strongly with the bolometric L$_{Bol}$ parameter from S11.  Furthermore, in order to obtain a uniform measurement of the Eddington fraction, we used the BH mass based on \MgII \ (VO09). Thus, rather than using the Eddington fraction listed among the measurements of S11, we used the following equation for the Eddington luminosity for all of our quasars:
\begin{equation}
L_{\textrm{Edd}} = 3.2 \times 10^{4} \left( \frac{M_{\textrm{MgII}}}{M_{\sun}} \right) L_{\sun}
\end{equation}
Throughout the remainder of this paper, when referring to the Eddington fraction we mean $L/L_{Edd}$ where $L$ is the bolometric luminosity (as provided in S11), and $L_{Edd}$ is as defined in the above equation 2 (not necessarily the same as provided in S11).   The luminosity L used on its own will generally refer to the continuum luminosity at C IV: $L_{cont,C IV}$.

Figure ~\ref{fig:Lz} shows the distribution of our sample in log luminosity vs. redshift space, where the luminosity is the continuum luminosity at \CIV\ ($L_{cont,CIV}$) as defined in equation 1.  Then in Fig.   ~\ref{fig:Eddz}, and Fig.  ~\ref{fig:Mz} we show the logarithmic distribution of Eddington fraction and BH mass (based on \MgII \ VO09) vs. redshift.  Our physical parameters each span roughly an order of magnitude for our sample.  We note that the range of these parameters in the quasar population is larger, extending to much smaller values.  Restricting the range as we do via this sample selection will create a bias that can drive differences in quasar properties to smaller values than might be found using the full population of quasars, and in that sense our results will be likely to represent lower limits.  Future surveys going to deeper flux limits with higher SNR would therefore be welcome.

\subsection{Spectral principal component analysis (SPCA)}
We use SPCA to deconstruct quasar spectra into a number of spectral principal components (SPCs), a subset of which can be added together, along with the mean spectrum, to produce spectra largely devoid of noise and absorption features.  
The SPCA weights can also be used to quantify quasar similarity.

We follow the approach of Francis et al. (1992), who first performed SPCA on the ultraviolet spectra of a large sample of quasars. We normalized all of the spectra to unit total integrated flux in our wavelength range, which focuses only on the variation in spectral shape rather than flux differences. For the principal component analysis itself, we used python's PCA task which generates a set of spectral principal components along with a vector of coefficients for each spectrum. We kept the first 25 principal components as these are sufficient to explain about 1$\sigma$ (67.5\%) of the variance and keeping more would only have served to fit the noise better rather than real physical features.

\begin{table}
  \centering
  \caption{Percentage of variance explained by each principal component in the sample.}
  \label{table:variance}
  \begin{tabular}{cc}
    \hline
Principal Component & \% variance explained \\
\hline
 1  & 29.1178\\
 2  & 19.6919\\
 3  & 3.2258 \\
 4  & 2.7731 \\
 5  & 2.0857 \\
 6  & 1.7300 \\
 7  & 1.4366 \\
 8  & 1.0334 \\
 9  & 0.8065 \\
10  & 0.6370 \\
11  & 0.5353 \\
12  & 0.4570 \\
13  & 0.4333 \\
14  & 0.3998 \\
15  & 0.3625 \\
16  & 0.3477 \\
17  & 0.3078 \\
18  & 0.2992 \\
19  & 0.2886 \\
20  & 0.2741 \\
21  & 0.2656 \\
22  & 0.2494 \\
23  & 0.2419 \\
24  & 0.2346 \\
25  & 0.2269 \\
sum & 67.4615\\
\hline
  \end{tabular}
\raggedright{

The second column shows the percentage of the variance explained by the first 25 principal components used on the sample. Those components are numbered in the first column. The last line is the sum of all the percentages of the variance explained by these 25 principal components. Thus it shows that keeping the first 25 principal components explains 67.4615 \% of the variance in the sample.
   }

\end{table}

\begin{figure}
\centering
\hspace{0cm}
   \includegraphics[width=9.5cm,trim={1cm 0 0 0},clip]{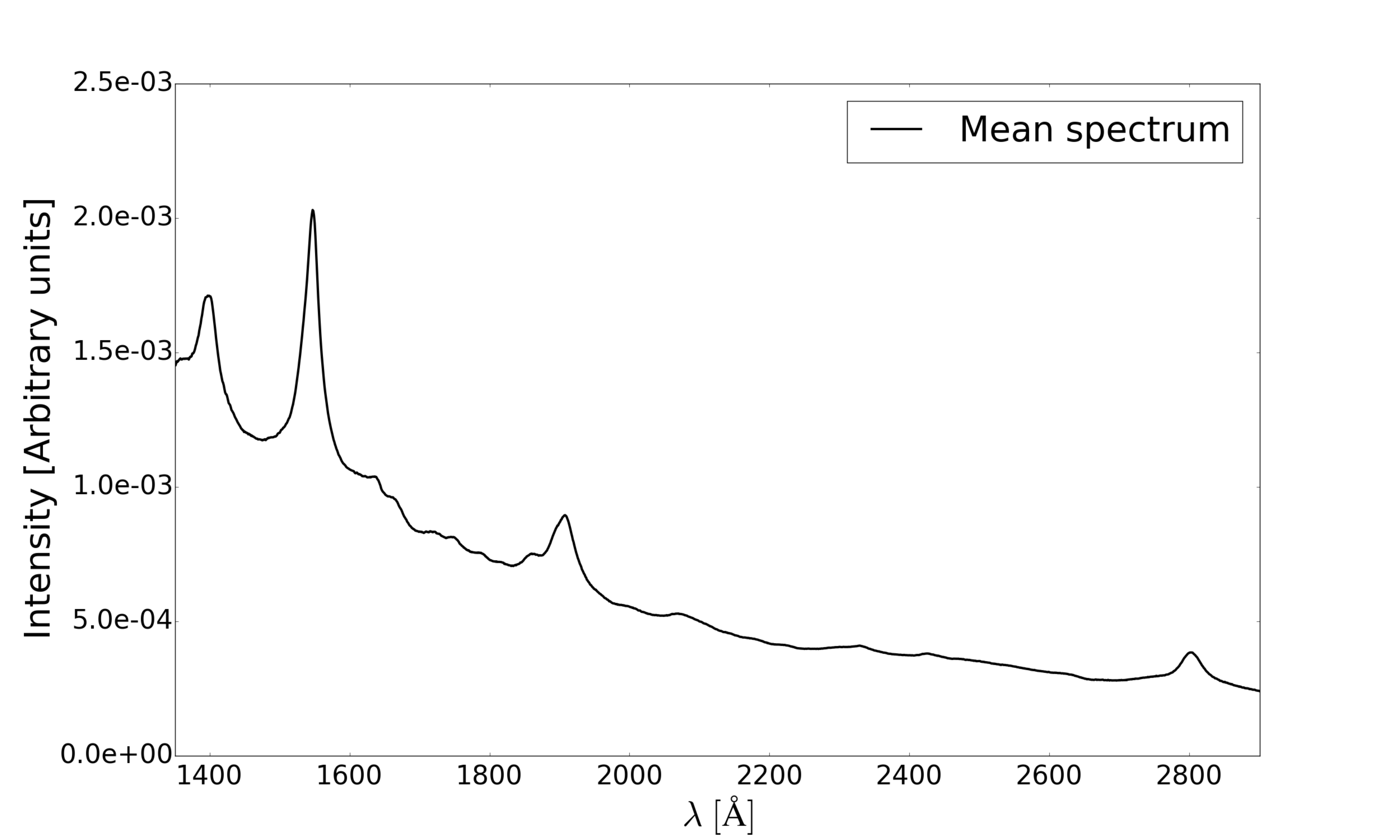}
   \includegraphics[width=9.5cm,trim={1cm 0 0 0},clip]{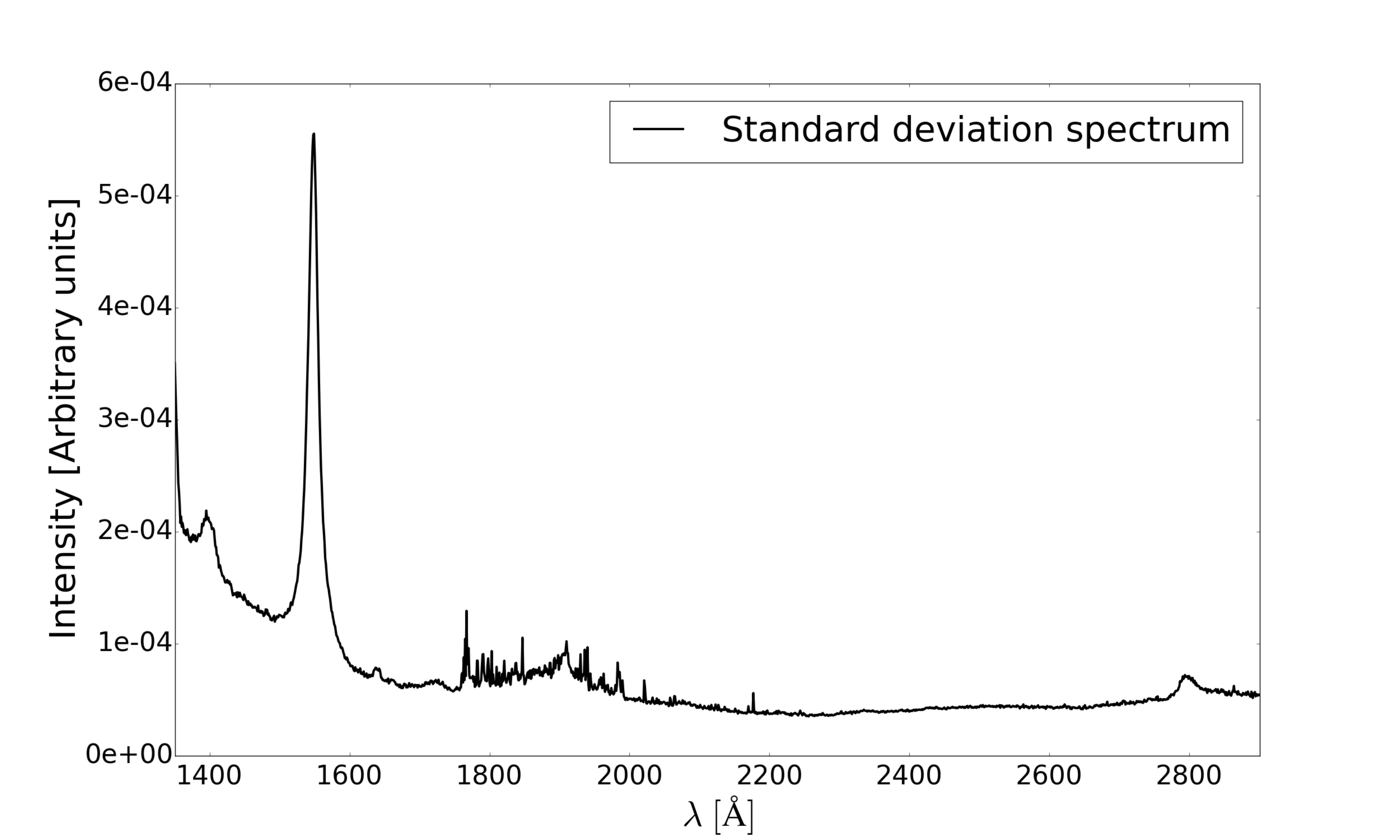}
   \vspace{0cm}
  \caption{Mean (top) and standard deviation (bottom) spectra.  The integrated mean spectrum is normalized to unity.   From left to right, the four most prominent emission lines are the $\lambda$1400 feature, C IV, C III], and Mg II.  We note that the spikes at wavelengths around 1900 \AA\ in the standard deviation spectrum result from poorly subtracted O I $\lambda$5577 sky lines in the original SDSS spectra.}
  \label{fig:MeanSTD}
\end{figure}

Figure ~\ref{fig:MeanSTD} shows the mean and standard deviation spectra for our full sample.  We note some spikes in the standard deviation spectrum that have an origin associated with the O I $\lambda$5577 sky line in individual spectra with similar redshifts.
A careful examination of these first 25 SPCs fails to uncover any sky line spikes.  Similarly we do not find any narrow C IV absorption doublets in these SPCs.  Given more than 5000 quasar spectra and a range of redshifts as well as absorption velocities, none of these features appear in the SPCs we use for spectral reconstruction.

\begin{figure}
\centering
\hspace{0cm}
   \includegraphics[width=9.5cm,trim={3cm 0 0 0},clip]{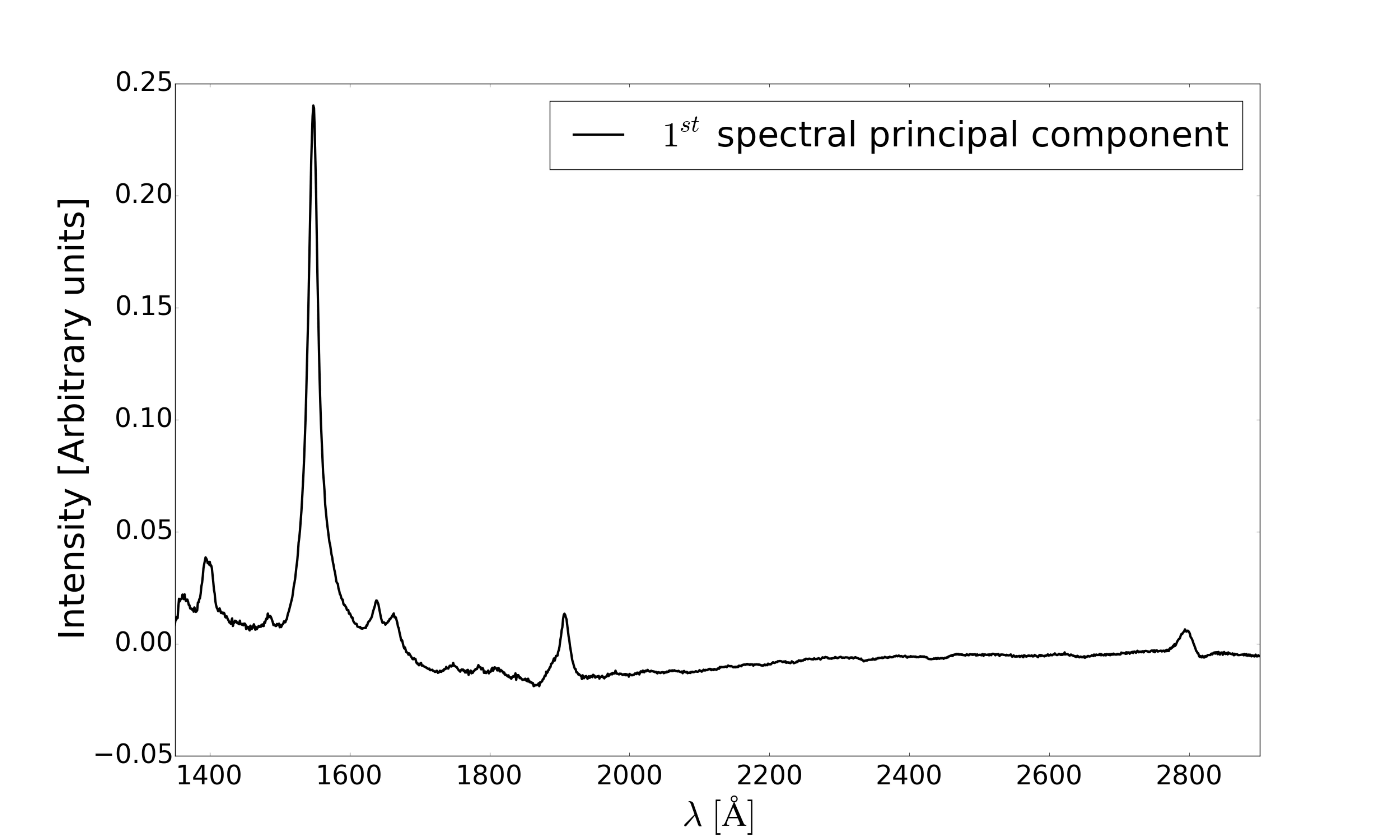}
   \includegraphics[width=9.5cm,trim={3cm 0 0 0},clip]{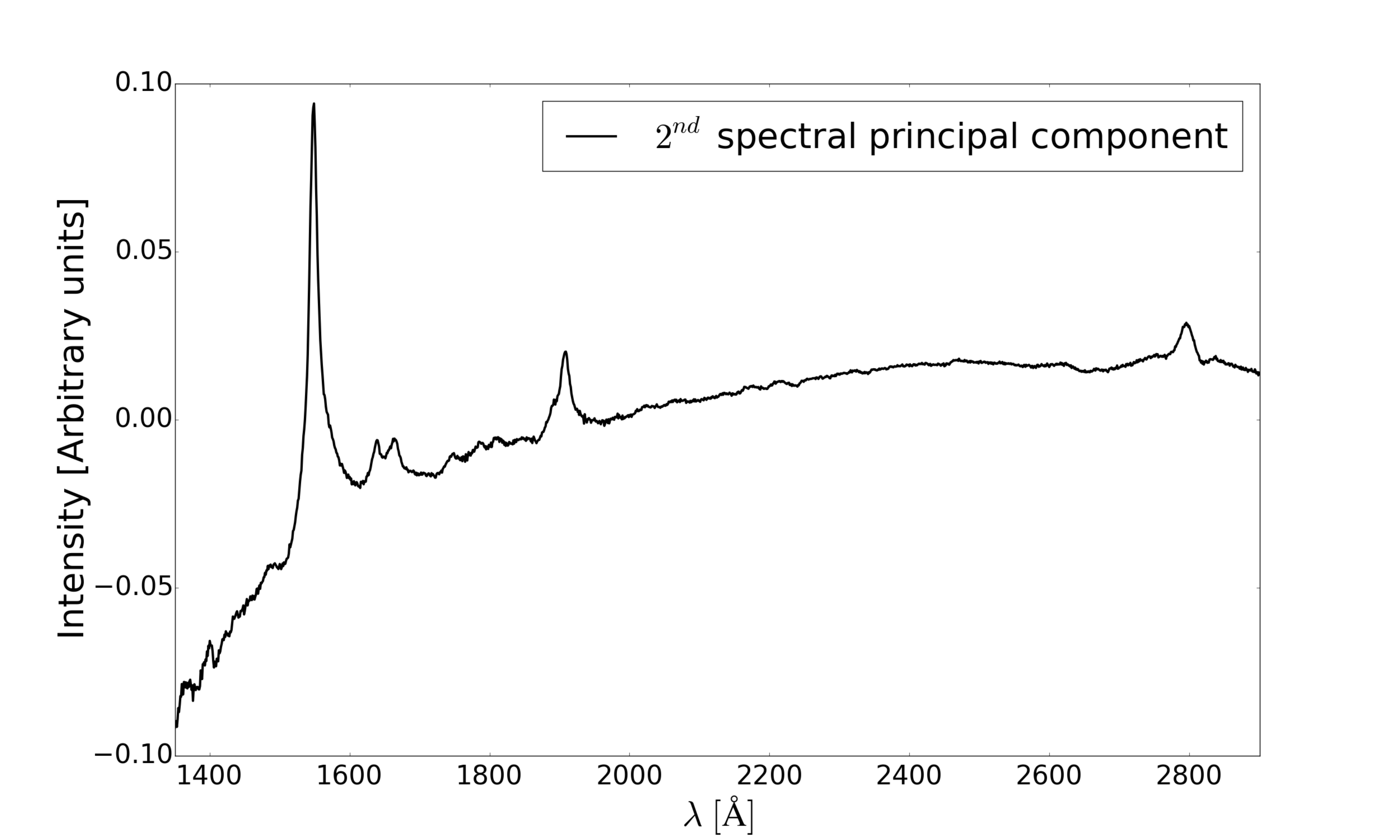}
   \vspace{0cm}
  \caption{First two spectral principal components, accounting for 29\% and 20\% of the spectral variance, respectively.  }
  \label{fig:PCAs}
\end{figure}

Table ~\ref{table:variance} provides the variance associated with the first 25 SPCs.  Figure ~\ref{fig:PCAs} plots the first two SPCs, which together account for 48.8\% of the variance.  SPC1 has emission lines narrower than the mean spectrum, and is roughly consistent with both the SPC1s of Francis et al. (1992) and Shang et al. (2003), and the Intermediate Line Region (ILR) spectrum of Brotherton et al. (1994) (see also P{\^a}ris et al. 2011).  SPC2 is dominated by continuum variation suggestive of dust reddening, along with some associated emission-line variation.  The third SPC and higher each contribute to only a very small part of the variance and we do not plot them.

We used the SPCA to reconstruct individual spectra according to the equation:
\begin{equation}
R = \sum_{i=1}^{25} SPC_i \cdot c_i
\end{equation}
Where $R$ is the reconstructed spectrum vector, $SPC_i$ is the $i^{th}$ spectral principal component vector, and $c_i$ is the multiplicative coefficient for the spectrum being reconstructed, corresponding to the $i^{th}$ principal component.

\begin{figure}
\centering
\hspace{0cm}
   \includegraphics[width=9.5cm,trim={1cm 0 0 0},clip]{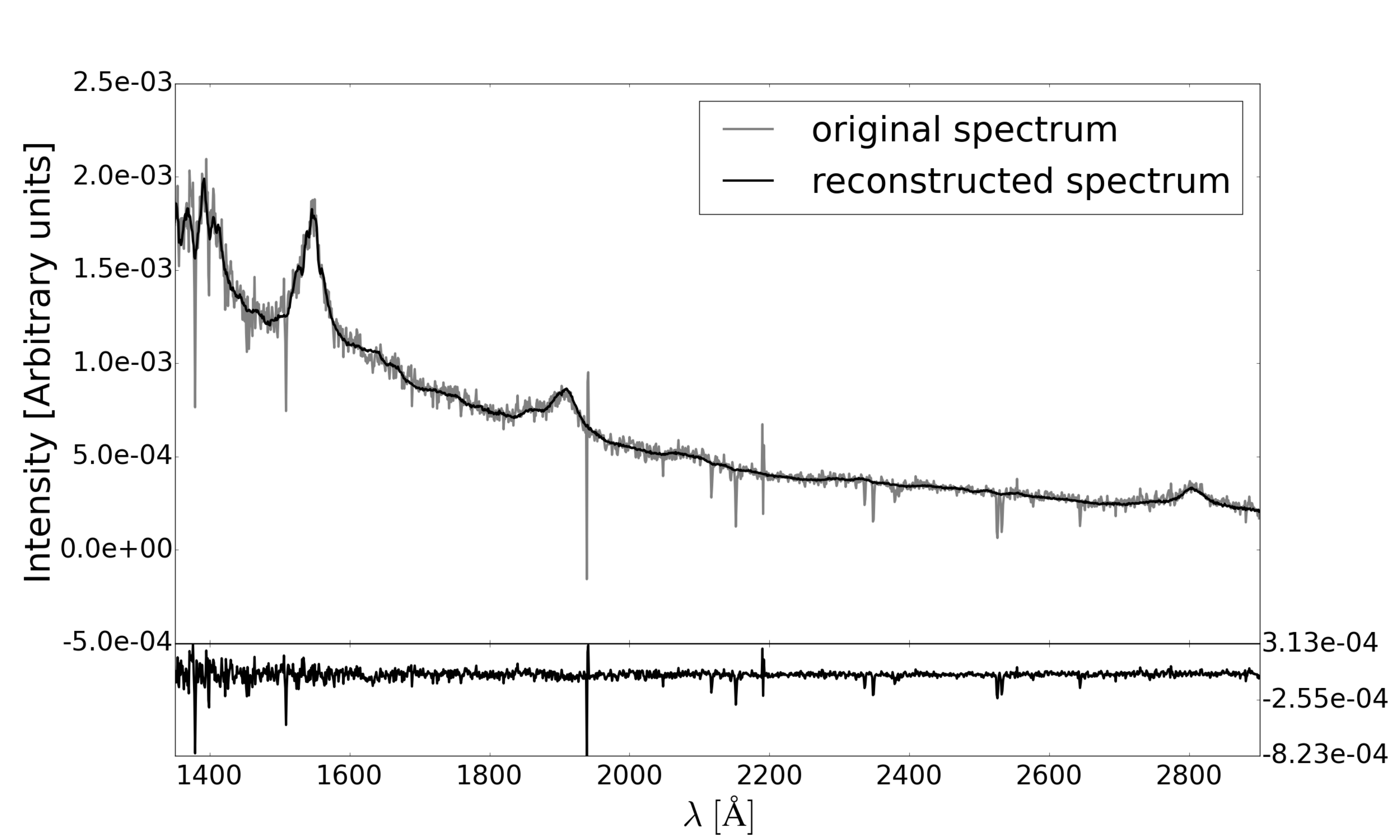}
   \includegraphics[width=9.5cm,trim={1cm 0 0 0},clip]{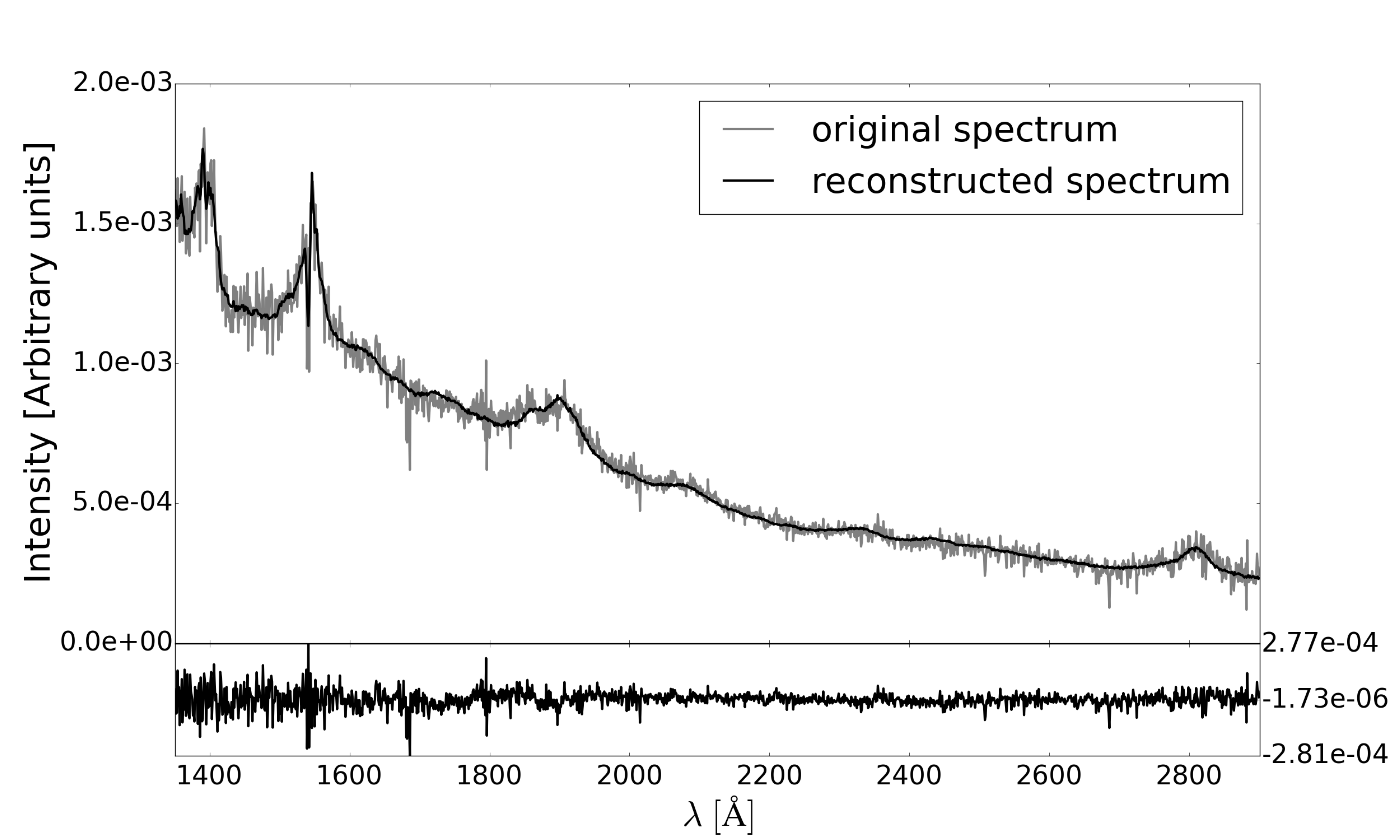}
   \vspace{0cm}
  \caption{Top panels: two reconstructed spectra. The original spectra are plotted in grey while the principal component reconstruction is in black. Bottom panels: the residual of the reconstructed spectrum subtracted from the original spectrum}
  \label{fig:Reconstructed}
\end{figure}

Figure ~\ref{fig:Reconstructed} shows two randomly selected principal component reconstructions using the first 25 spectral component spectra (black) plotted over the original spectra (grey). In the bottom panels we show the result of subtracting the reconstructed spectrum from the original spectrum. This figure  and more extensive visual inspection indicates the success of our approach.

\subsection{Hyperspace distance criterion}
\begin{figure*}
\centering
\hspace{0cm}
   \includegraphics[width=8.5cm,trim={3cm 0 0 0},clip]{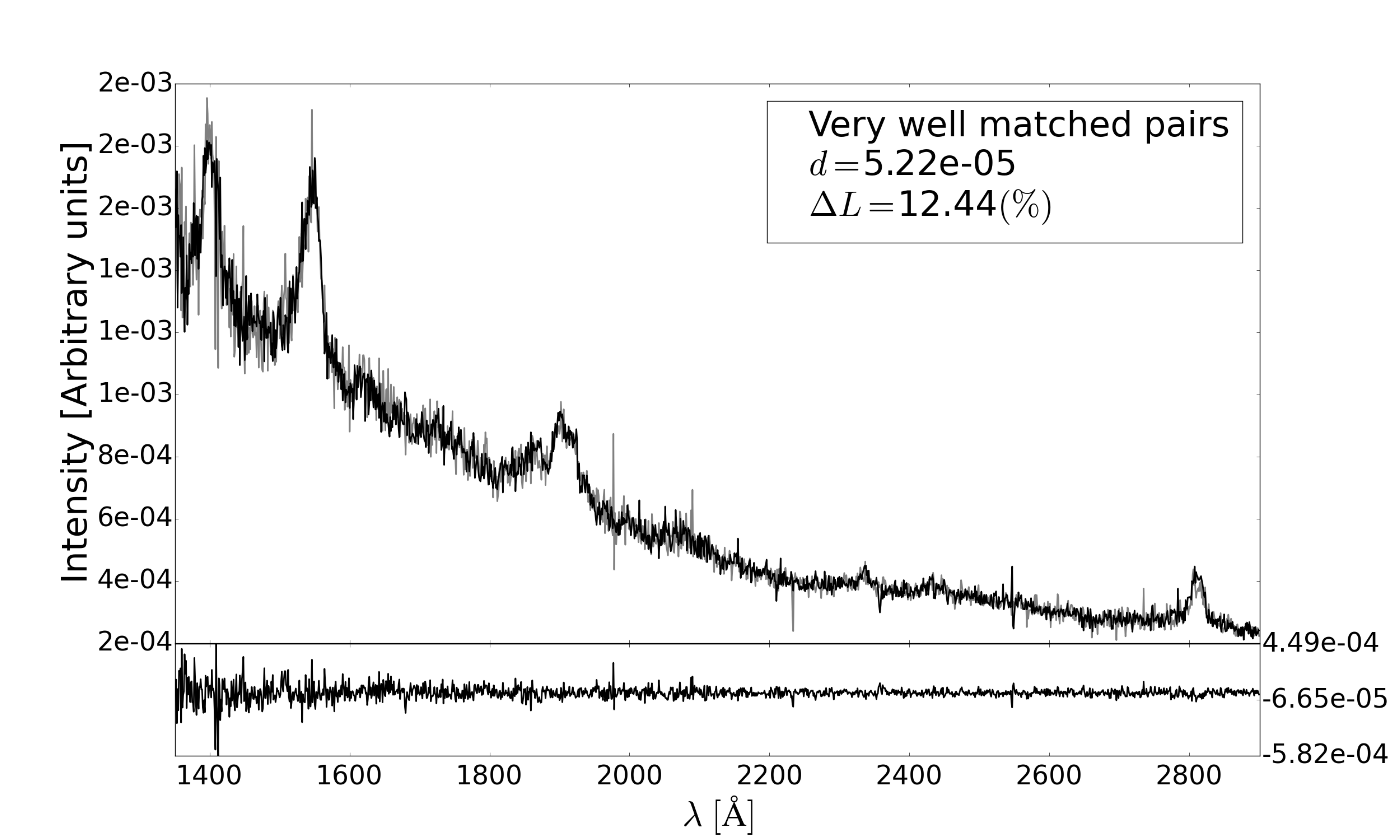}
   \includegraphics[width=8.5cm,trim={3cm 0 0 0},clip]{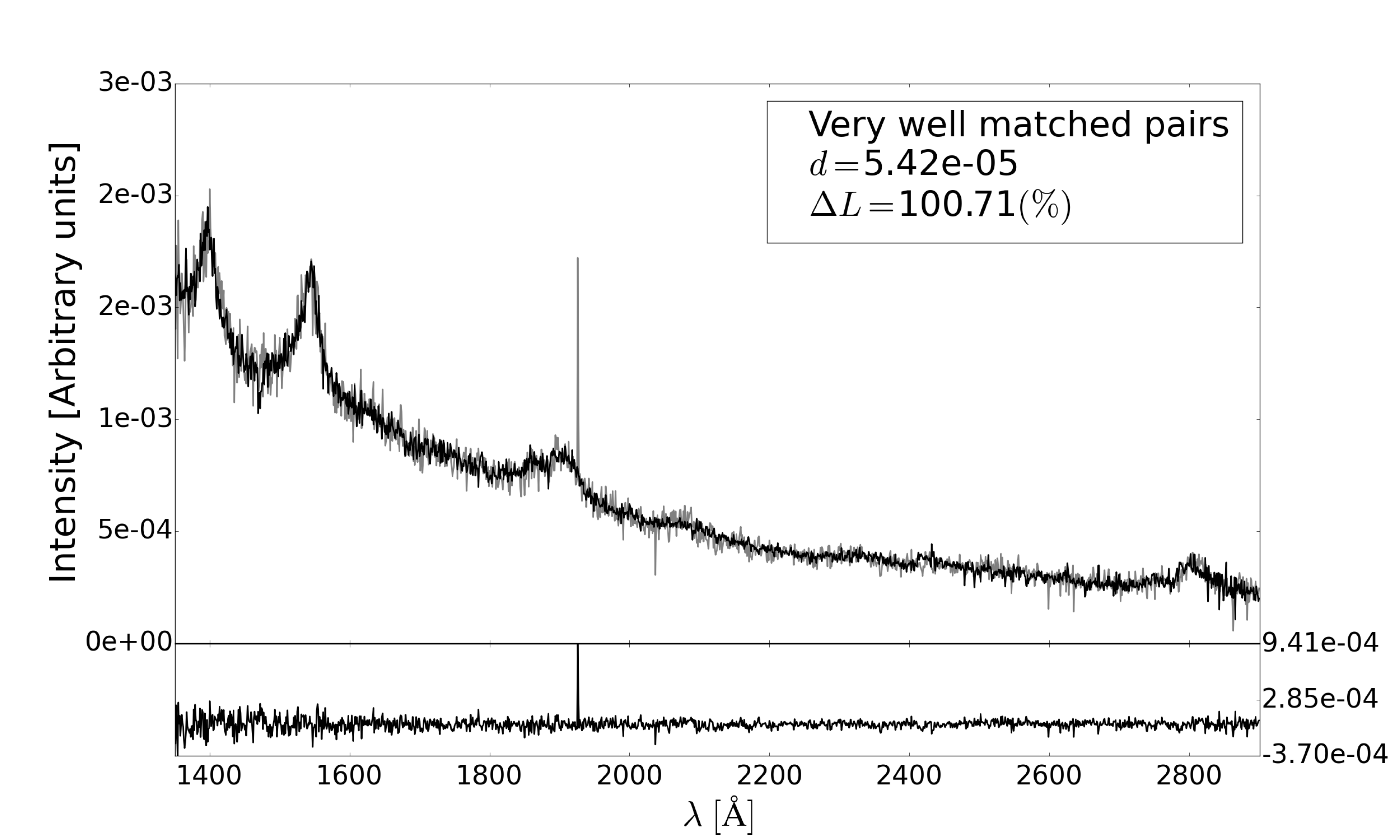}
   \includegraphics[width=8.5cm,trim={3cm 0 0 0},clip]{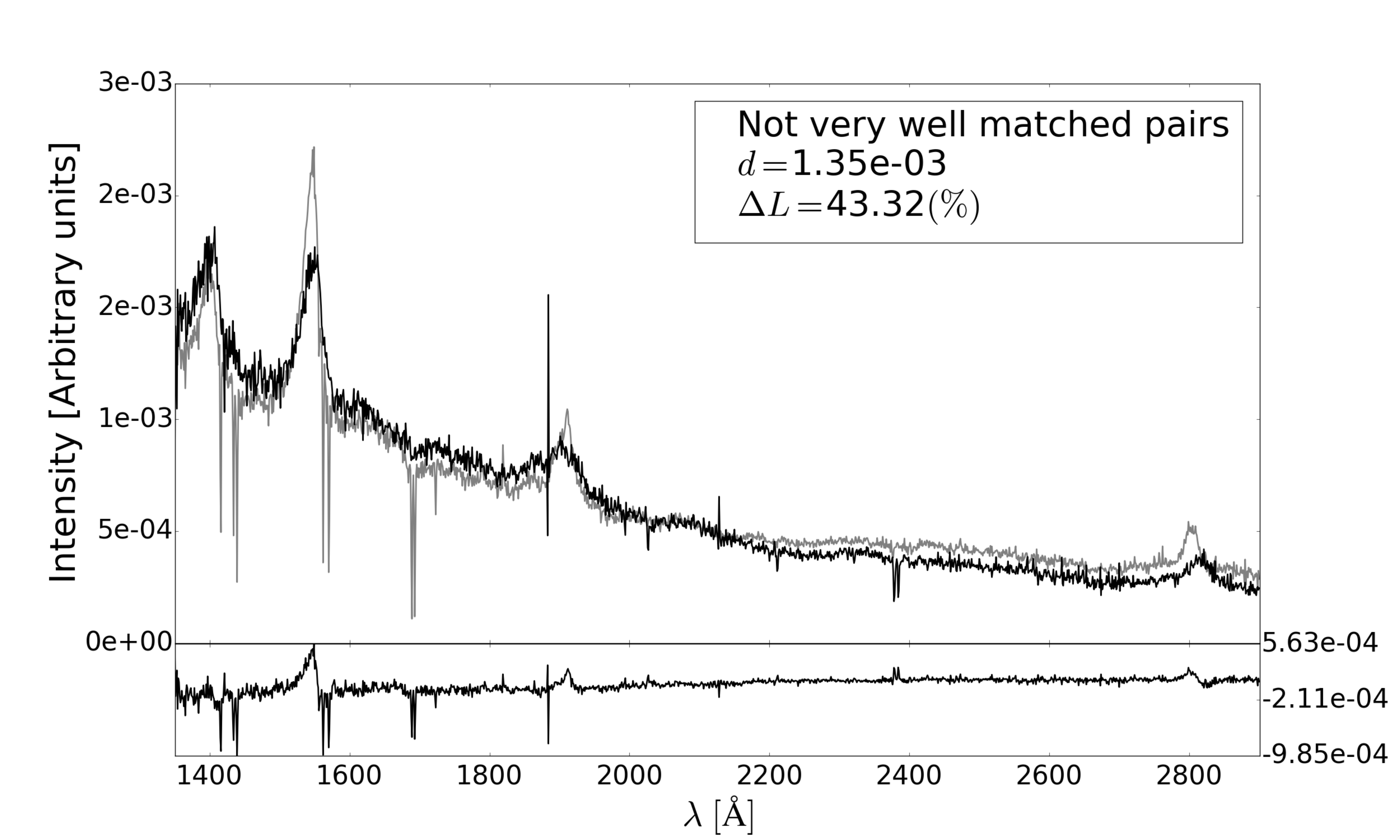}
   \includegraphics[width=8.5cm,trim={3cm 0 0 0},clip]{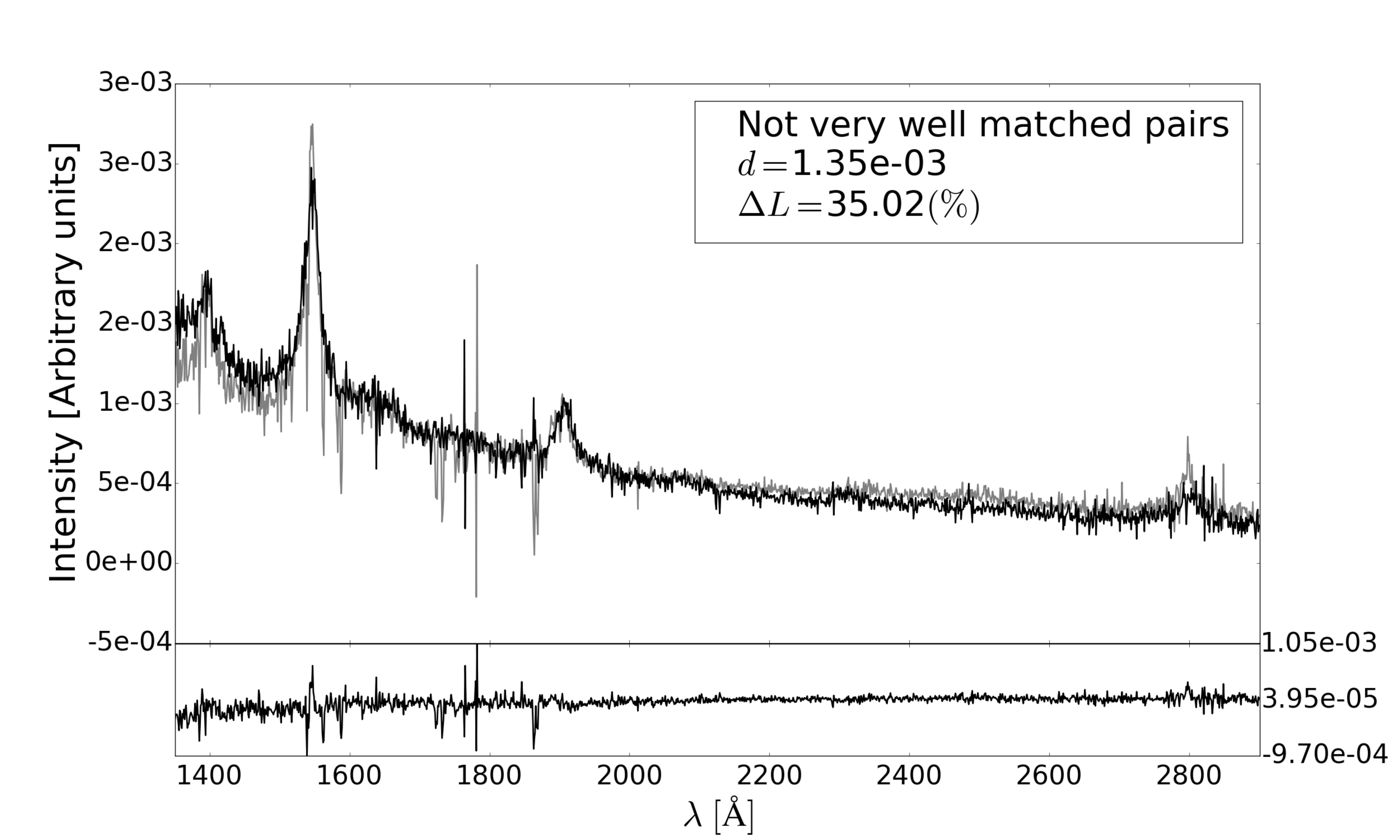}
   \includegraphics[width=8.5cm,trim={3cm 0 0 0},clip]{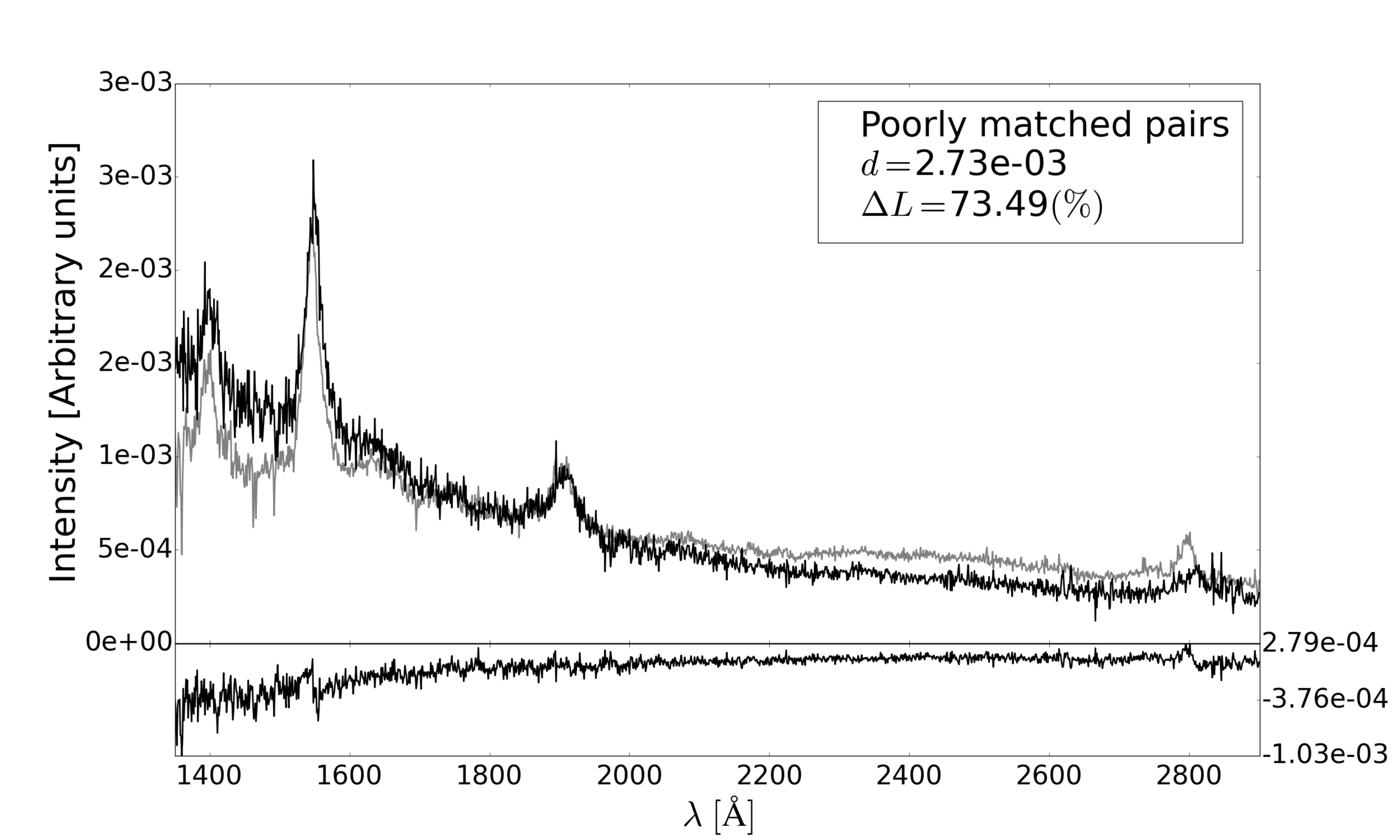}
   \includegraphics[width=8.5cm,trim={3cm 0 0 0},clip]{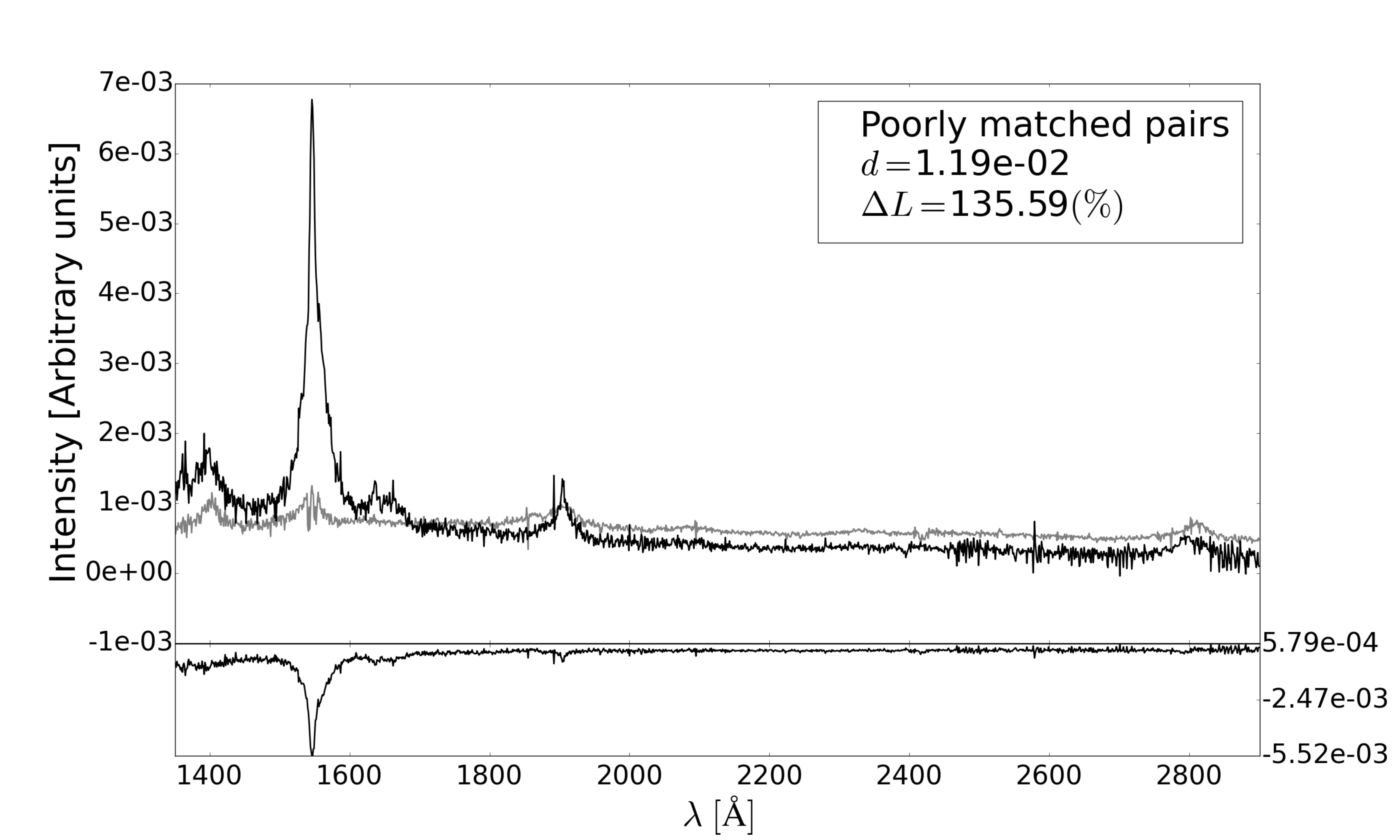}
   \vspace{0cm}
  \caption{Six pairs of spectra (grey and black) in increasing hyperdistance separation. As $d$ increases, the spectra become poor matches, but for individual spectral pairs, $\Delta L$ does not necessarily increase.}
  \label{fig:Pairs}
\end{figure*}

The hyperspace distance between two spectra is defined as the weighted norm between their coefficient vectors:
\begin{equation}
d = \sqrt{\sum_{i=1}^{25}w_i\cdot (c_{1,i}-c_{2,i})^2}
\end{equation}
The parameter $w_i$ is the variance associated with the $i^{th}$ principal component, and $c_{1,i}, c_{2,i}$ are the multiplicative coefficient weights for the  $i^{th}$ principal component of the first and second spectra. The smaller $d$\ is for a given pair, the more similar two spectra are. Figure ~\ref{fig:Pairs} shows several paired spectra, ranging from our best matches to poorly matched pairs, according to our `hyperspace distance' criterion. These figures illustrate the effectiveness of our criterion since spectra separated by a small $d$ are good matches, while those separated by larger $d$ are poor matches. Furthermore, the percentage difference in $L_{cont,CIV}$ is given, showing that for individual pairs, we can have very well matching spectra with both well matched luminosities as well as more poorly matched luminosities.  We explore this further in the next section.  Also, as a remark, we note that these `hyperspace distances' have arbitrary but relative units, since their values strongly depend on how the spectra were normalized. All that matters in our analysis are the relative distances.

\begin{table*}
 \centering
 \begin{minipage}{180mm}
  \caption{The best SDSS DR7 spectral doppelgangers.  The full table is available online.}
  \begin{tabular}{@{}cccccccccccc@{}}
  \hline
SDSS Name & Plate & Fiber & MJD & SDSS Name & Plate & Fiber & MJD & Hyper Distance d & $\Delta M_{BH}$ & $\Delta L_{cont,CIV}$ & $\Delta L/L_{Edd}$ \\
& & & & & & & & & $(\%)$ & $(\%)$ & $(\%)$ \\
\hline

125805.31+512207.0 & 0886 & 020 & 52381 & 105523.56+414214.7 & 1362 & 109 & 53050 & 5.21678E-05 & 64.40 & 12.90 & 46.78\\
082810.35+324134.5 & 0932 & 473 & 52620 & 161406.54+203757.6 & 2206 & 453 & 53795 & 5.42040E-05 & 93.45 & 99.91 & 17.00\\
110939.35+464349.7 & 1438 & 466 & 53054 &  135424.90+243006.3 & 2119 & 281 & 53792 & 5.50692E-05 & 6.91 & 16.86 & 24.97\\
102820.33+561204.6 & 0947 & 139 & 52411 & 154948.71+263047.8 & 1654 & 258 & 53498 & 5.51142E-05 & 11.50 & 24.29 & 35.76\\
133700.08+024925.5 & 0528 & 489 & 52022 & 153647.69+152149.9 & 2782 & 485 & 54592 & 5.56260E-05 & 95.24 & 30.74 & 65.84\\
010523.24+140930.3 & 0421 & 035 & 51821 & 141404.80+134455.8 & 1706 & 468 & 53442 & 5.86475E-05 & 13.79 & 71.57 & 58.56\\
120324.32+072049.3 & 1623 & 192 & 53089 & 152848.98+200649.1 & 2164 & 380 & 53886 & 6.05314E-05 & 72.53 & 59.52 & 21.33\\
172209.44+273042.9 & 0980 & 147 & 52431 & 154640.75+300123.7 & 1390 & 372 & 53142 & 6.10330E-05 & 82.31 & 74.62 & 3.45\\
130717.34+125206.1 & 1696 & 485 & 53116 & 133220.98+161158.5 & 2606 & 101 & 54154 & 6.11398E-05 & 98.77	 & 37.27 & 70.92\\
111040.10+101304.7 & 1221 & 562 & 52751 & 111949.84+225143.1 & 2493 & 007 & 54115 & 6.21757E-05 & 86.10	 & 8.45 & 81.35\\

\hline
\end{tabular}
\end{minipage}
\end{table*}

We calculated hyperspace distances for the full pair-wise comparison of every spectrum in our sample, a total of 15,415,128 pairs.
Table 2 provides the 1272 pairs with the smallest hyperdistance separations. Those correspond to the ones with hyperdistance from $5.13\times10^{-5}$ (our best match) to $1.04\times10^{-4}$ (roughly twice the best match distance).  All appear to be excellent matches by visual inspection, which is not consistently true moving into larger separations.  We also provide the differences in continuum luminosity, Eddington fraction, and black hole mass for each pair.  These may be useful and of interest for follow-up studies.

\section{Analysis and results}

\subsection{doppelgangers}

\begin{figure}
\centering
\hspace{0cm}
   \includegraphics[width=9.5cm]{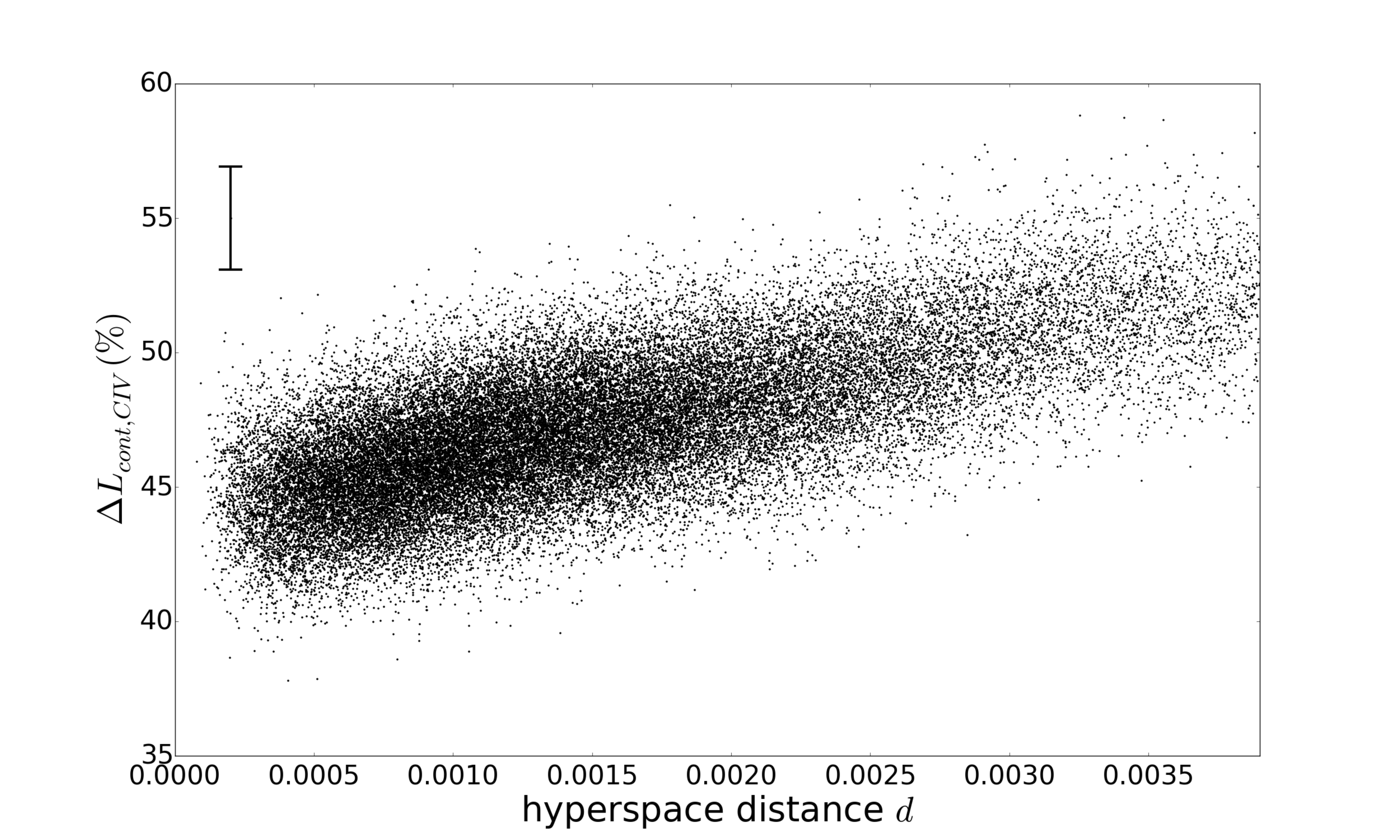}
   \vspace{0cm}
  \caption{$\Delta L$ vs. hyperspace distance, with 347 pairs of quasars per bin. The error bar represents a typical standard error for each bin.}
  \label{fig:deltaL}
\end{figure}

\begin{figure}
\centering
\hspace{0cm}
   \includegraphics[width=9.5cm]{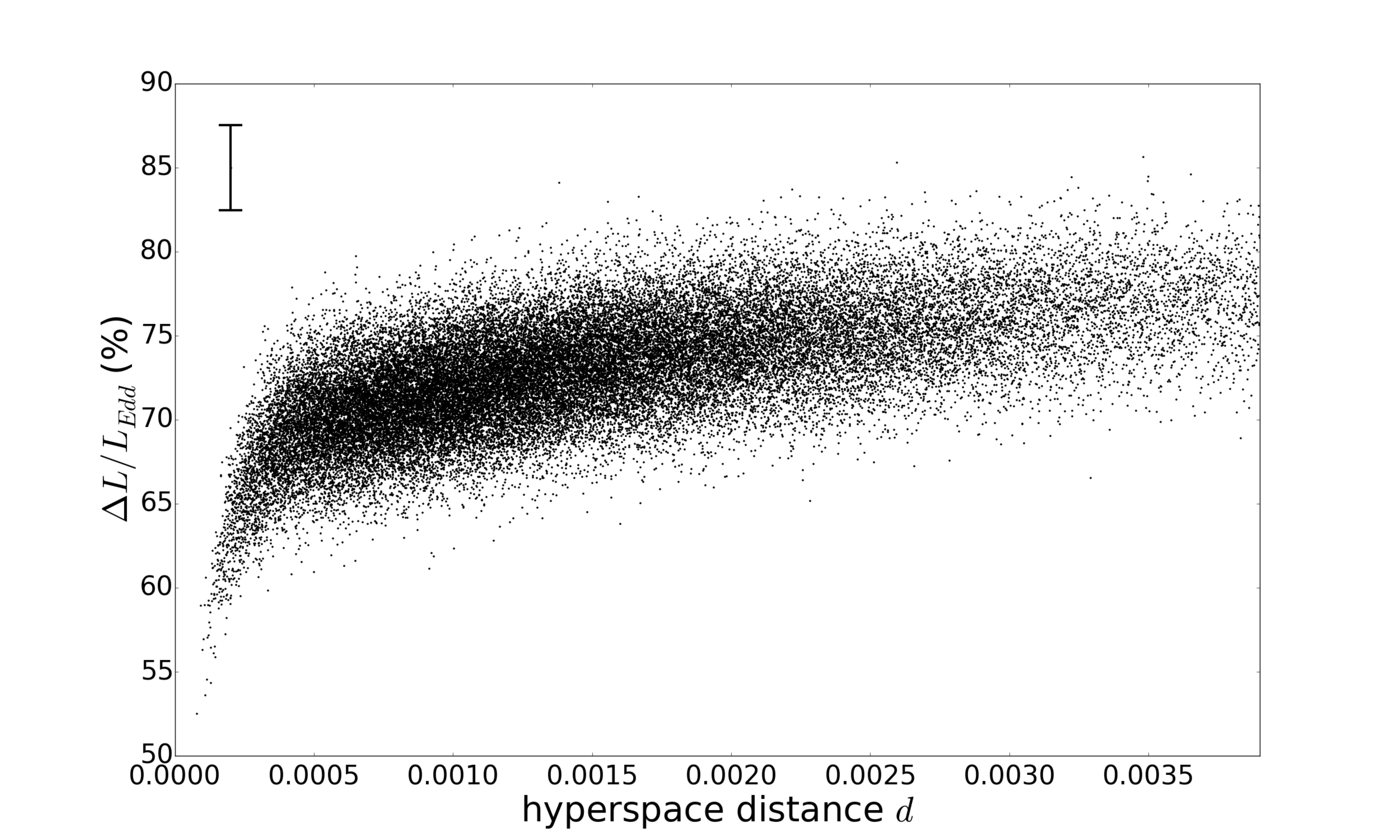}
   \vspace{0cm}
  \caption{$\Delta L/L_{Edd}$ vs. hyperspace distance, with 347 pairs of quasars per bin. The error bar represents a typical standard error for each bin.}
  \label{fig:deltaEddMgII}
\end{figure}

\begin{figure}
\centering
\hspace{0cm}
   \includegraphics[width=9.5cm]{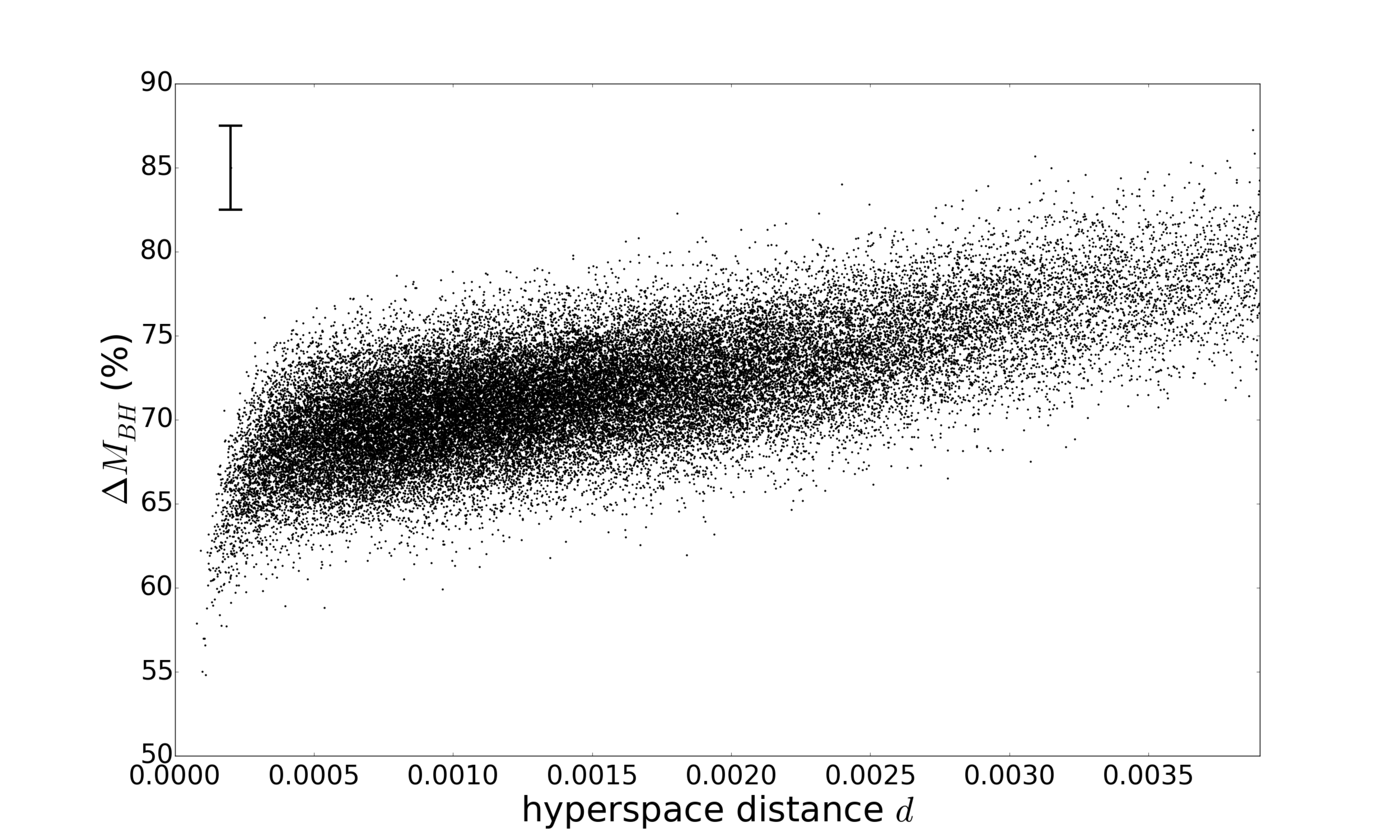}
   \vspace{0cm}
  \caption{$\Delta M_{BH}$ vs. hyperspace distance, with 347 pairs of quasars per bin. The error bar represents a typical standard error for each bin.}
  \label{fig:deltaM}
\end{figure}

We compute differences in quasar properties by taking the absolute value of the raw differences and normalizing by the average of the values of each quasar pair.
Figures ~\ref{fig:deltaL},~\ref{fig:deltaEddMgII}, and ~\ref{fig:deltaM} show the difference in the continuum luminosity $\Delta L$, the difference in the Eddington fraction $\Delta L/L_{Edd}$, and the difference in the Mg II black hole mass $\Delta M_{BH}$ vs. the hyperspace distance $d$. We restrict the x-axis range to $d<0.0039$, leaving the 42,843 best matches out of the total (the best 0.279\%); the differences between randomly selected pairs of quasars are significantly larger than the scale on these figures and we focus on the transition between doppelgangers and worse matches, thus avoiding the more random distribution of the very poorly matched pairs. Furthermore, the data were binned into groups of 347, and only the mean value of those bins is plotted. In the top left corner of figures ~\ref{fig:deltaL},~\ref{fig:deltaEddMgII}, and ~\ref{fig:deltaM} we plot a typical standard error in the mean. This standard error is computed as the standard deviation in each bin divided by the square-root of the number of pairs per bin (347): $\sigma/\sqrt{N}$. For Figure ~\ref{fig:deltaL}, showing $\Delta L$ vs. $d$, we have $\sigma/\sqrt{N} = 1.9150 \pm$ 0.0969, for Figure ~\ref{fig:deltaEddMgII}, showing $\Delta L/L_{Edd}$ vs. $d$, we have $\sigma/\sqrt{N} = 2.5230 \pm 0.0948$, and for Figure ~\ref{fig:deltaM} showing $\Delta M_{BH}$ vs. $d$, we have $\sigma/\sqrt{N} = 2.4984 \pm 0.0922$.


As we can see, there is a clear correlation between $\Delta L$, $\Delta L/L_{Edd}$, $\Delta M_{BH}$, and the hyperspace distance $d$. Note that we have zoomed in the most similar spectra,
and given the large contribution of SPC1 and SPC2, these must be similar in all these well-matched
pairs.  The differences are much larger for dissimilar spectra.
Having more similar ultraviolet spectra does indeed lead to more similar quasar properties.
The means and standard deviations for our best matched bins are: a $45.95\pm 32.83\%$ difference in luminosity, a $52.51\pm 35.99\%$ difference in Eddington ratio, and a $57.85\pm 39.55$\% difference in mass. Considering the continuum luminosity, the corresponding difference in magnitude is $0.50$ mag. 
In order to get the very smallest hyperspace distances, higher-order SPCs must also agree,
which allow the emission-line profiles to match in detail -- this accounts for the slope change 
in Figures 9 and 10.  The FWHM Mg II will have intrinsically the same profiles in the best
doppelgangers, but the measurement errors in FWHM Mg II from S11 are on average 17\%.
The calculated black hole masses are proportional to
(FWHM Mg II)$^2$ L$^{0.5}$, and $L/L_{Edd}$ ratios are proportional to (FWHM Mg II)$^{-2}$ L$^{0.5}$.  Therefore, given the large range in luminosities observed for doppelgangers, it 
makes sense that the observed differences in $M_{BH}$ and $L/L_{Edd}$ are similar but
somewhat larger than those for luminosity given the additional measurement errors.
While on average the differences in luminosity, Eddington ratio, and mass are on order of 50\% for the best doppelgangers, the scatter is nearly as large. The differences between individual quasars, even for the best matching doppelgangers, can be significant, and is primarily driven by 
luminosity variation.

\begin{figure}
\centering
\hspace{0cm}
   \includegraphics[width=8.5cm]{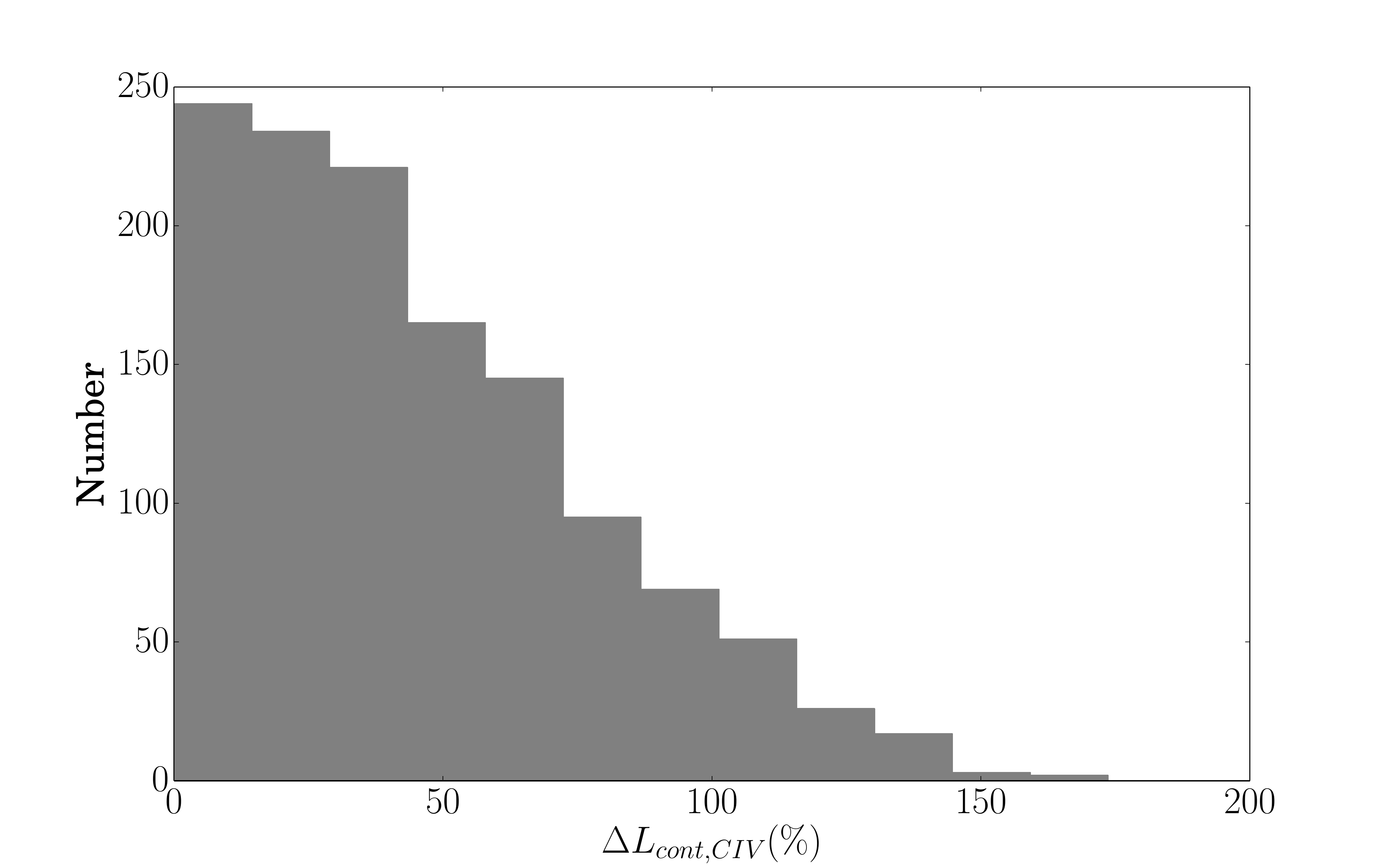}
   \vspace{0cm}
  \caption{The distribution of $\Delta L_{cont,CIV}$ for the best-matched 1272 doppelgangers from Table 2.}
  \label{fig:deltaLdop}
\end{figure}
\begin{figure}
\centering
\hspace{0cm}
   \includegraphics[width=8.5cm]{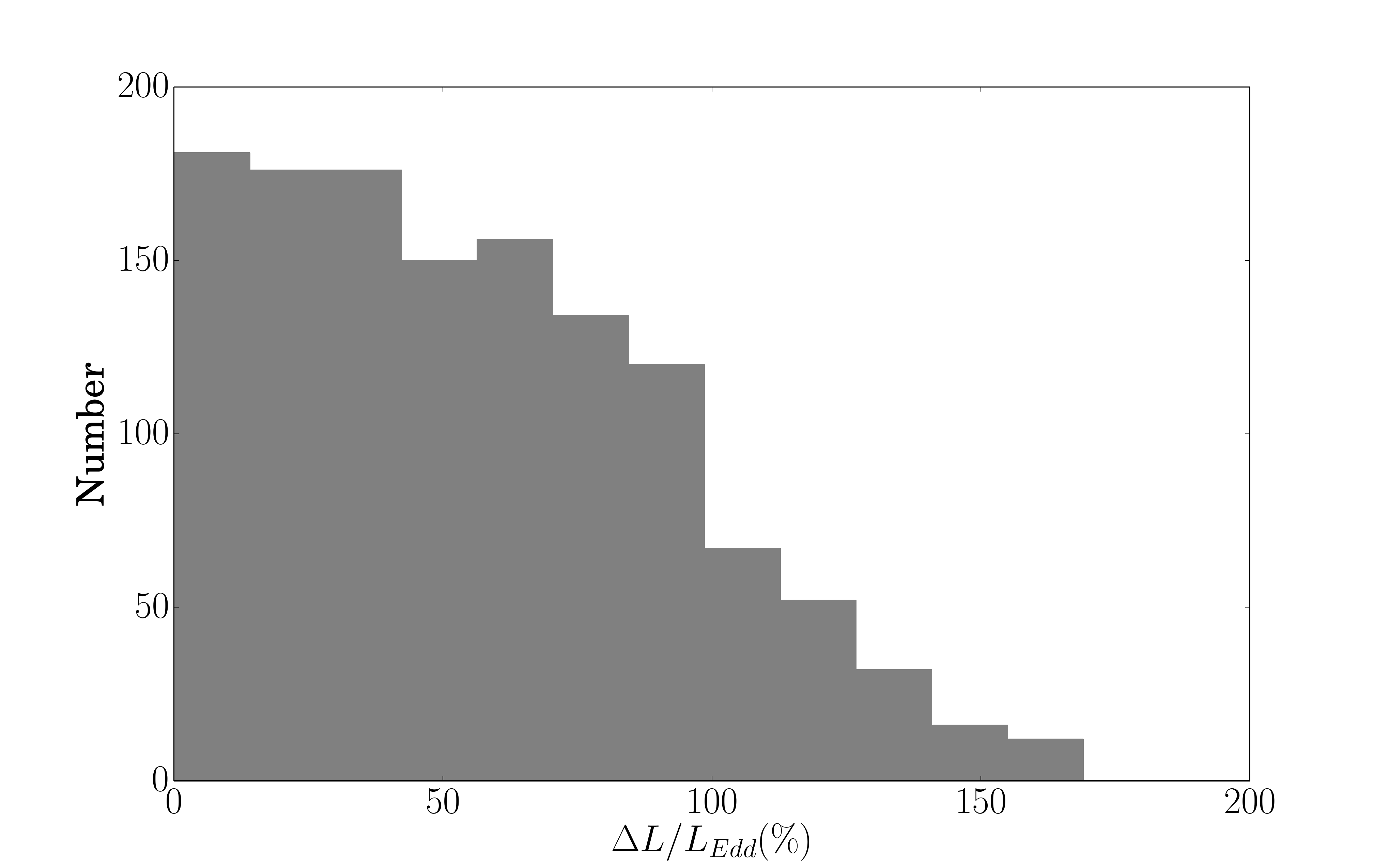}
   \vspace{0cm}
  \caption{The distribution of $\Delta L/L_{Edd}$ for the best-matched 1272 doppelgangers from Table 2.}
  \label{fig:deltaEddMgIIdop}
\end{figure}
\begin{figure}
\centering
\hspace{0cm}
   \includegraphics[width=8.5cm]{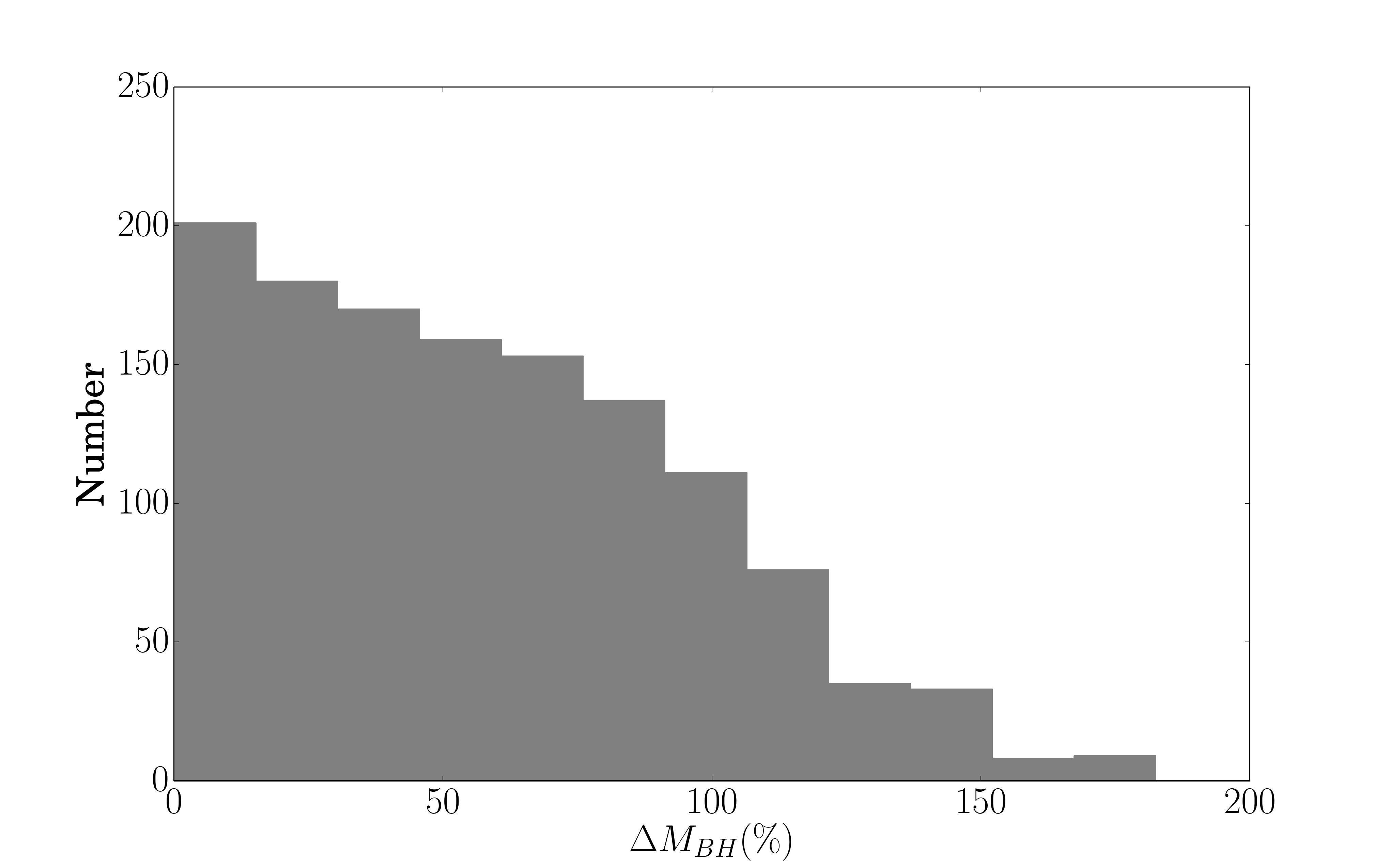}
   \vspace{0cm}
  \caption{The distribution of  $\Delta M_{BH}$ for the best-matched 1272 doppelgangers from Table 2.}
  \label{fig:deltaMdop}
\end{figure}

Figures 11, 12, and 13 show histograms of the differences for our best-matched doppelgangers from Table 2.  While the average is one valid way of characterizing the differences in the doppelganger properties, it is more characteristic for comparisons to use 1$\sigma$ variations.  We can provide these by calculating the standard deviations of the distributions with the addition of a reflection around x = 0.  Doing this, we determine that the standard deviations for  $\Delta L_{cont,CIV}$ is 56.6\%,
for $\Delta L/L_{Edd}$ is 67.7\%, and for $\Delta M_{BH}$ is 70.3\%.  Considering the continuum luminosity difference standard deviation, the corresponding difference in magnitude is $0.63$ mag, or about 0.19 dex, which
is the most appropriate figure of merit to compare against most other predictors of luminosity quoting uncertainties. 

\subsection{The case of high Eddington ratio}
\begin{figure}
\centering
\hspace{0cm}
   \includegraphics[width=9.5cm]{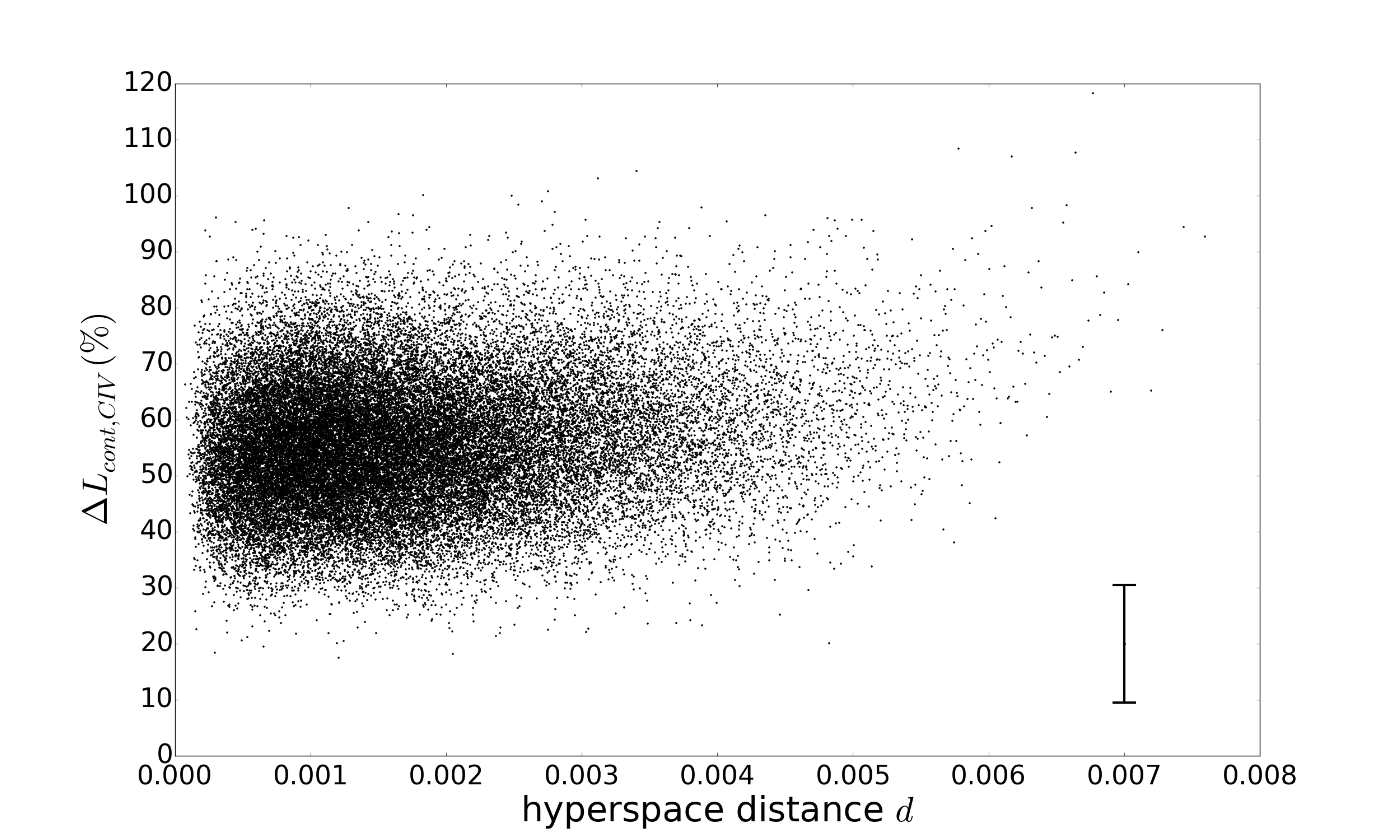}
   \vspace{0cm}
  \caption{$\Delta L$ (for $L/L_{Edd} > 0.9$) vs. hyperspace distance, with 13 pairs of quasars per bin.  The error bar shown represents a typical standard error for each bin.}
  \label{fig:deltaLhighEdd}
\end{figure}

Marziani \& Sulentic (2014) suggested that quasars with high Eddington ratio should be better candidates for matching luminosities. We therefore considered separately a subsample of the 1062 quasars with the highest  Eddington ratios ($L/L_{Edd} > 0.9$).  Again, we binned our data, but this time using only 13 pairs per bin, in order to keep the same ratio of pairs per bin over total number of pairs. This is shown in Figure ~\ref{fig:deltaLhighEdd}, the bottom right corner has a typical standard error $\sigma/\sqrt{N} = 10.51\pm2.10$. On average, our best matched bin reaches a $66.26\pm 38.15\%$ difference in continuum luminosity. If instead we look at the best 100 pairs (rather than the best 13) we get the following average: $58.79\pm 40.07\%$ difference in luminosity. And again, if we look at the best 347 pairs (just like we did with the full sample) we get the following average: $55.03\% \pm 37.54\%$. This later value is not a clear improvement over the full sample, and therefore this analysis shows no evidence of improvement when reducing the sample to high Eddington ratio quasars.  To restate this point, while it may be the case that the average pair of high Eddington-fraction pairs is better matched in luminosity than any average pair, there is no improvement compared to pairs that have well-matched spectra, as determined in moderately high signal-to-noise ratio data, in general.

We note that, for our sample, if we evaluate the scatter around the Baldwin effect in the luminosity direction, we get a very comparable standard deviation of about 0.20 dex or 58\%, albeit with a tail to much higher values (factors of many) than what is present in Figure 11 (factors of a several).  The standard deviation is not necessarily the only figure of merit to be considered, as we will discuss below.

\subsection{Correlation Analysis}

Table ~\ref{table:spearman} shows the Spearman rank correlation coefficients for several parameters of our sample, primarily our quasar properties and measurements of the C IV line, as well as our first two SPC weights. For such a large sample even relatively small correlation coefficients result in a very low probability of arising by chance.   All the correlations in the table have probabilities p $< 10^{-4}$) except for these 10: $M_{BH}$ with $EW_{CIV}$ has p = 0.745, $M_{BH}$ with $V_{off,C_{IV}}$ has p = 0.16815, $SPC\ 1$ with $L/L_{Edd}$ has p = 0.890, $SPC\ 2$ with $M_{BH}$ has p = 0.0260, $SPC\ 2$ with $SPC\ 1$ has p = 0.761, $z$ with $EW_{CIV}$ has p = 0.0329, $z$ with $FWHM_{CIV}$ has p = 0.0019, $z$ with $M_{BH}$ has p = 0.0075, $z$ with $L/L_{Edd}$ has p=0.0204, and $z$ with $SPC\ 2$ has p = 0.0014. So except for $M_{BH}$ with $EW_{CIV}$, $M_{BH}$ with $V_{off,C_{IV}}$, $SPC 1$\ weight with $L/L_{Edd}$, and $SPC\ 2$ with $SPC\ 1$, the Spearman rank probabilities are all statistically significant.

\begin{table*}
  \centering
  \caption{Spearman rank correlation coefficients}
  \label{table:spearman}
  \begin{tabular}{lcccccccccc}
    \hline
&$L_{cont,CIV}$&$L_{bol}$&$EW_{CIV}$&$FWHM_{CIV}$&$M_{BH}$&$L/L_{Edd}$&$\alpha_{CIV}$&$V_{off,C_{IV}}$&$SPC\ 1$&$SPC\ 2$\\
    \hline
$L_{bol}$&\textbf{0.9571}\\
$EW_{CIV}$&-0.3307&-0.2794\\
$FWHM_{CIV}$&0.1617&0.1209&-0.3896\\
$M_{MgII}$&0.3047&0.2885&-0.0044&0.2408\\
$L/L_{Edd}$&0.3088&0.3494&-0.1767&-0.1454&\textbf{-0.7463}\\
$\alpha_{CIV}$&-0.0962&-0.1322&-0.0599&0.2152&0.0913&-0.1690\\
$V_{off,C_{IV}}$&0.1177&0.1254&-0.1698&0.3392&0.0185&0.0575&0.0758\\
$SPC\ 1$&-0.2496&-0.2265&\textbf{0.7165}&\textbf{-0.5063}&-0.1570&-0.0019&-0.4731&-0.1978\\
$SPC\ 2$&-0.2281&-0.1713&0.4855&-0.2574&0.0299&-0.1332&\textbf{0.5119}&-0.1243&-0.0041\\
$z$&0.1912&0.2442&-0.0286&0.0417&0.0359&0.1215&0.1112&0.3483&-0.1021&0.0428\\
\hline
  \end{tabular}
\raggedright{

Spearman rank correlation coefficients for several properties of our sample. The correlation coefficients greater than 0.5 (in absolute value) are shown in bold.
   }
\end{table*}

Recognizing that most of the correlations are formally quite significant, few of the correlation coefficients are large.  We have placed the only five correlation coefficients larger than 0.5 in bold.  The strongest correlation is between $L_{cont,CIV}$ and $L_{Bol}$,
which are both based on continuum measurements at similar but not identical wavelengths.  The variation in the Eddington fraction is strongly driven, although not totally driven, by the variance in the black hole mass.  The EW and FHWM of C IV are strongly correlated with SPC1, as has been long known 
(e.g., Brotherton et al. 1994b).  Finally, the spectral index of the C IV continuum is quite strongly correlated with SPC2, consistent with the second SPC having a strong continuum slope.

Of interest are the correlations relating SPC1 and SPC2 to the quantities involved in the Baldwin effect and the modified Baldwin effect.  SPC1 is dominated by a C IV line core, low-velocity gas producing emission narrower than the mean C IV profile, and referred to as the Intermediate Line Region, and it is the variation of this component that leads to the inverse correlation between EW and FWHM in C IV, while having little effect on 
other lines like $\lambda$1400 feature (e.g., Wills et al. 1993).  Also related, although at a weaker level, is the blueshift of the C IV line (e.g., Brotherton et al. 1994a; Richards et al. 2011).  SPC1 has the strongest correlation with EW CIV, and is inversely correlated with luminosity in a way that contributes significantly to the Baldwin effect (previously noted by Shang et al. 2003).  Moreover, SPC2, which also possesses a strong narrow C IV component correlated with a reddened continuum, also contributes to the Baldwin effect.  Formally, the variance in luminosity associated
with SPC1 and SPC2 together matches that associated with EW C IV; SPC1 and SPC2 together
can quantitatively account for the entire Baldwin effect, which is only about 11\% of the variance in luminosity.  There is indeed significant scatter in the correlation.

Perhaps surprisingly, given that the modified Baldwin effect has been seen in many samples to be stronger than the Baldwin effect itself -- although not in the present sample -- the correlation coefficient between the Eddington fraction and SPC1 (representing the largest source of spectral variation in the sample) is the smallest in the entire table.  We will explore these findings in more detail in the discussion below.

\section{Discussion}

\subsection{doppelgangers}

In many fields of science there are ``lumpers'' and ``splitters.''   That is, there are scientists who look at a class of things, and either focus on the similarities, or on the differences, and quasar astronomy is no exception.  Over the past several decades, many papers have noted the general similarity of quasar spectra, and some have tried to explain the selection effects leading to this general similarity, e.g., the ``locally optimally emitting clouds'' picture of Baldwin et al. (1995) (see also Korista et al. 1997).  Others have focused on exploring and explaining the variation in quasar spectra, e.g., Boroson \& Green (1992) and Shen \& Ho (2014).

Quasar spectra do vary in systematic ways from object to object, but with the large numbers of objects now available as the result of surveys like the SDSS, it is possible to find doppelgangers with very similar ultraviolet spectra.  We think it is interesting to identify these, as this does not seem to have been done quantitatively and in such a detailed manner as we have done.  Often in astronomy the problem is to use our knowledge of physics to interpret the physical properties of an object by modeling the observed spectrum.  Here we are inverting that question.  We can ask, what are the limits of the physical properties that can be determined from a spectrum, based on the observation that two nearly identical spectra do not always seem to have the same luminosity (or other parameters)?  Beyond this current project it will be interesting to see whether or not the differences we note here correlate with differences at other wavebands (e.g., the X-ray) and that more carefully matched doppelgangers using more observations may result in better agreement in properties.

\subsection{Quasar Parameter Prediction, Uncertainties, and Limitations}

Quasar spectra certainly depend on the properties of the quasars.  Trying to isolate and predict a single property from an ultraviolet spectrum has associated uncertainties and limitations, however.  What are the origins of the differences in properties for quasar doppelgangers?

Now in the era of precision cosmology, especially compared to past decades when quasar properties were first studied, it is unlikely that fluxes are converted to luminosities with large errors.  Additionally, our redshift range is relatively small: $1.8 \leq z \leq 2.17$.  The luminosity distance for our cosmology only varies from 44.5 Gly to 56.0 Gly over these redshifts.  If we change the cosmology to a slightly different, more recent preferred value, $\Omega_{\Lambda}$ = 0.696,  $\Omega_{m}$ = 0.286, and H$_{0}$=69.6 km s$^{-1}$ Mpc$^{-1}$ (Bennett et al. 2014), the luminosity distances only change to 45.3 Gly and 57.0 Gly, a $\sim$3-4\% difference when squared as required to convert fluxes to luminosities.  This is negligible in comparison to other issues, assuming no major error in cosmology.

The next practical consideration concerns measurement errors, which we previously mentioned
in \S 3.1.  According to S11, continuum flux is measured to about 1\%, and FWHM Mg II 
to an average value of 17\%; the latter measurement is squared when used to compute a black hole mass, so there would be a corresponding $\sim$34\% uncertainty in mass on average. Very small differences in hyperspace distance, as in the case of our doppelgangers, will match line profiles extremely well, although there would still be systematic errors to measure absolute FWHM Mg II from spectral reconstructions.   

It is worthwhile to consider the plots of hyperspace distance vs. $\Delta L_{cont,CIV}$, $\Delta M_{BH}$, and $L/L_{Edd}$ in more detail.  In the case of luminosity, there is a correlation with an  approximately linear slope from small values to larger values over the range plotted.  In the case of the other two, which depend upon measurements of the Mg II line in detail and which has relatively weak variation with SPC1 and SPC2, there is a sharp change in slope at relatively small hyperspace distances.  That is indicative of the need for the higher-order spectral components to get the Mg II profile right in enough detail to get a good match, which only happens in the matches with the smallest hyperspace distances.

There are other issues to consider other than just measurement uncertainties.  When ``quasar properties'' are discussed, the phrase generally refers to the true (bolometric) luminosity, the true Eddington ratio, and the true black hole mass.  There are significant uncertainties in the determination of all of these, and the uncertainties based on imperfect measurements contribute to some of the differences we observe.  That is, FWHM Mg II and a flux/luminosity measurement is an imperfect predictor of black hole mass (see, e.g., Shen 2013) even if all the information about the mass is inherent in the ultraviolet spectrum -- which may not be true.

Moreover, quasars are axisymmetric sources, with the ultraviolet continuum thought likely to be emitted by an accretion disk (e.g., Shields 1978; Sun \& Malkan 1983).  The continuum emission emitted from a disc is anisotropic (e.g., Netzer 1985; Jackson et al. 1989; Nemmen \& Brotherton 2010; Runnoe et al. 2013).
If the line emission is emitted with a different degree of isotropy than a disc, there will be variation in their EWs associated with orientation, although only marginal effects have been found in C IV (e.g., Runnoe et al 2014).  The C IV profile itself does show an orientation dependence (e.g., Vestergaard 2000; Runnoe et al. 2014) as
does the Mg II profile (Runnoe et al. 2013a).  Thus, while orientation creates some uncertainty between the line-of-sight flux and the bolometric luminosity more generally, there is information in the spectra and 
doppelgangers likely ought to share similar orientation.

There exist some additional, more fundamental issues.  There are some factors that are in essence free parameters when using photoionization modeling to fit quasar spectra.  One of the most notable is the covering fraction, which is the fraction of the sky covered by broad-line region clouds as seen from the accretion disc.  This parameter needs to be set at $\sim$10\% -- or a few percent for the narrow line region (NLR) and ILR, to a few tens of percent for the BLR more generally (Brotherton et al. 1994b).  This is a parameter that can likely vary from object to object despite similar underlying properties.  Similarly the column density can likely vary within gas in the BLR and from object to object, resulting in a range of emission-line strengths with ionizing flux (e.g., Korista et al. 1997).  

Another phenomenon that can lead to variation between the emission-line spectrum and the continuum is variability coupled with the relatively large size of the BLR compared.  Kaspi et al. (2007), for instance, find variation timescales of several years or longer for the C IV BLR in high-luminosity quasars.  On the timescale of several years, for luminous quasars at high redshifts like those in our sample, De Vries et al. (2005) find that the structure function is about $-$0.5 to $-0.6$ magnitudes in the log, or approximately 30\%.  This means that emission lines need not track the continuum exactly, and exactly what the relative response is will have an effect on the observed spectral properties like the line EWs.  A variation of $\sim 30\%$ would account for much, but not quite all ($\sim 45\%$), of the luminosity variation we see in the doppelgangers.

There may be yet more issues, but the issues above alone are likely to explain the differences in quasar properties for objects with otherwise identical ultraviolet spectra.  Still, there remains hope to use quasars as standard candles, or candles that can be calibrated well enough, for cosmological investigations.  At high redshifts, the number of observable quasars still vastly outnumbers the number of supernovas.

\subsection{Implications for Cosmology}

Folatelli et al. (2013) report the uncertainty in predictable supernova luminosities to be 0.15 mag or about 15\%.  Watson et al. (2011) have also proposed that the dispersion in the BLR radius-Luminosity relation is a similar 0.13 dex.  Clustering of quasars has also been used as a cosmological standard ruler, relying on Baryon Acoustic Oscillations (BAOs), and has proven an effective cosmological probe (e.g., Anderson et al. 2012).

What can quasars as standard candles do?

The traditional problem with using relationships like the Baldwin effect to predict quasar luminosities is that the scatter is too large.  What does that mean?  It means, despite some large numbers present in modern surveys, that selection effects create biases.  That is, in the presence of a flux limit, the large scatter in combination with a range in redshift, creates an artificial correlation.  

What needs to be possible to achieve this the selection of quasars at one redshift, with a luminosity predictor, that can be compared in an unbiased way to a sample at a different redshift.  This will allow an unbiased comparison of the two samples.  The first problem is the large scatter of the Baldwin effect that places some potential matches below the flux limit, biasing the comparison.  The second problem is that the flux limit is too high in many surveys.

Following a standard procedure (e.g., Wampler et al. 1984), quasars can be used as standard candles if a sample selected at one redshift can be compared in an unbiased way to quasars at another redshift, in large enough numbers, to test cosmological models.  This is challenging currently, but potentially plausible in the near future, given quasar numbers and depths of future surveys (e.g., DESI, Font-Ribera et al. 2014).  SPCA has provided one way to match quasars, perhaps at the cost of large numbers, but with increased precision.

There are a number of issues to consider quantitatively.  First, our doppelgangers indicate that the worst matches in luminosity correspond to $\Delta$L $\sim$ 150\%.  This corresponds to a factor of 7  between the luminosities of the pair.  The luminosity range in a sample must be able to accommodate a scatter this large or differences in luminosity limits (for flux-limited samples) will bias any comparisons at significantly different redshifts.  Our sample in this paper is on order of an order of magnitude, and likely too small.   Additionally, schemes like that of Wampler et al. (1984) are likely too simple.  The quasar luminosity function differs with redshift, and this should be taken into consideration when finding matches.  An appropriate analysis for testing cosmology with quasars as standard candles will likely be best done using a modeling approach that takes into account all these factors, but in principal, with a deep enough sample, it is feasible.

\subsection{SPC1}

The first SPC is very similar to that of Francis et al. (1992), being dominated by a C IV core component in emission.  This is also consistent with the Intermediate Line Region (ILR) of Wills et al. (1993), photoionization modeled by Brotherton et al. (1994b) to be at larger distances than the rest of the C IV emitting region.  Denney (2012) noted that C IV line cores tended not to vary in reverberation mapping studies, consistent with the modeling results.  Moreover, it is this line core that leads to scatter in the correlation between the FWHM of C IV and H$\beta$ lines, the former line being systematically broader in the variable component (e.g., Onken et al. 2003) but often narrower in single-epoch spectra.  The C IV line core component is correlated with the [O III] $\lambda$5007 narrow line (Brotherton \& Francis 1999; Sulentic et al. 2007), tying together the principal sources of variation in the ultraviolet and the optical.  Boroson (2002) found that the [O III] line variation was strongly inversely correlated with $L/L_{Edd}$.

Interestingly, SPC1 is {\em not} significantly correlated at all with $L/L_{Edd}$.  This is an issue of potential concern, as the C IV core component correlates with [O III] $\lambda$5007 (Brotherton \& Francis 1999), and [O III]  $\lambda$5007 correlates with $L/L_{Edd}$ (Boroson 2002).  It is not necessary for correlations to be commutative, given that correlations have scatter, but it is an issue to investigate in the future.

To first order, there is likely a disc-like component of the BLR that reverberates (e.g., Onken et al. 2004), and a low-velocity non-virial component that correlates with the NLR and does not (e.g., Denney 2012).  This is the SPC1 component or ILR component that dominates object-to-object variation.  Based on Table 3, it likely is a major contributor to the Baldwin effect (e.g., Shang et al. 2003), but SPC2 also contributes.

\subsection{SPC2} 

Given the nature of the strong wavelength dependence of SPC2 one natural hypothesis is that this source of variance is associated with internal dust reddening (e.g., Baron et al. 2016).  Baker \& Hunstead (1996) show an apparent orientation effect between dust reddening and emission line EWs.  The second principal component, SPC2, is reminiscent of that effect.  Like SPC1, SPC2 has C IV lines narrower than the mean, indicative of the ILR, and the NLR.  Our SPC2 has the effect of amplifying, given the similar signs, SPC1 effects, adding orientation and continuum shape variation to luminosity.  Quantitatively, the EW of C IV accounts for about 11\% of the luminosity variance in our sample, as does the combination of SPC1 and SPC2, which correlate very strongly with the C IV EW and account for $\sim$75\% of its variance.

\subsection{Additional SPCs}

We note that none of the additional SPCs we use in the spectral reconstructions past the second have correlation coefficients with luminosity close to as large as those of SPC1 and SPC2.  They all individually account for only a small amount of both the spectral variance and the luminosity variance.  Still, there are some correlations worth mentioning.

The third spectral principal component shows a strong asymmetry in the C IV line and does correlate rather strongly with the C IV velocity offset ($r = 0.485$) at a much higher level than SPC1 and SPC2.  The set of SPC3, SPC4, and SPC5 together have similarly moderate correlations ($r = 0.25$ to $r=0.30$) with the Mg II-based black hole mass, and represent components that can account for a changing broad Mg II FWHM.

\section{Conclusions}

There exist quasars with extremely similar ultraviolet spectra, and the more similar the spectra, the more similar on average the luminosity.  The similarity in luminosity given similar spectra with similar broad line velocity widths also results in similar Eddington fraction and black hole mass.  The similarity in luminosity only approaches 50\% in the mean, however, and can be as large as factors of several in individual pairings.  Our findings identify the quantitative limits of using quasars as standard candles based on ultraviolet spectra alone, as well as the limits of our current parameter estimation techniques.

Our spectral principal component analysis (SPCA) recovers the same first principal component seen in previous work, which primarily has strong effects on the C IV emission-line profile in the ultraviolet,  and furthermore appears to have a strong contribution to the Baldwin effect.  Our second principal component appears likely to be consistent with orientation effects and also contributes to the Baldwin effect.

This study shows that when quasars match well spectroscopically there is general agreement in their luminosities, i.e., quasar emission lines in certain cases do indeed show correlation with luminosity. The most well-matched quasars can be calibrated to predict luminosity and hence test luminosity distance. This supports previous arguments for the potential to use quasars for high-redshift cosmology, but that potential is not yet realized with current samples.
An approach such as ours, using quasar doppelgangers, appears promising for cosmological studies, approaching supernova precision.  Deeper flux limits at high redshift will be advantageous.

\section*{Acknowledgments}
We thank the University of Wyoming course Astronomy 5460 for bringing together this collaboration.  We also thank Ohad Shemmer and Adam Myers for useful comments.

\label{lastpage}

\end{document}